\documentclass[12pt]{iopart}

\usepackage{graphicx}
\usepackage{harvard}

\newcommand\commentout[1]{}
\newcommand{\constant}{{\mbox {\rm constant}}}
\newcommand{\dx}{{\mbox{\rm d}}}
\newcommand\eqref[1]{(\ref{#1})}
\newcommand{\smfrac}[2]{{\textstyle{\frac{#1}{#2}}}}
\def\half{\smfrac{1}{2}}

\newcommand{\Lemaitre}{Lema\^{\i}tre}

\begin{document}

\title[Hubble's law and the expanding universe]{Milestones of general
  relativity: Hubble's law (1929) and
the expansion of the universe}

\author{Malcolm A H MacCallum}

\address{School of Mathematical Sciences, Queen Mary University of
  London, London E1 4NS, U.K. \ead{m.a.h.maccallum@qmul.ac.uk}}

\begin{abstract}
Hubble's announcement of the magnitude-redshift relation \cite{Hub29}
brought about a major change in our understanding of the
Universe. After tracing the pre-history of Hubble's work, and the
hiatus in our understanding which his underestimate of distances led
to, this review focuses on the development and success of our
understanding of the expanding universe up to the present day, and the
part which General Relativity plays in that success.
\end{abstract}
\submitto{\CQG}

\maketitle

\section{Introduction}
In a seminar which I attended as a new graduate student in 1966-67, the late
Dennis Sciama said that when he himself had started work in cosmology
in 1950 there was only one known fact about the Universe -- and that
fact later turned out to be wrong! The fact was that the Universe was
expanding, which was known from Hubble's law relating the
magnitudes and redshifts of galaxies, and what was wrong with it was
the expansion rate. Hubble's data led, in the simplest interpretation,
to a Universe younger than the geologically-known age of the Earth.

Redshift $z$ can be interpreted as due to the Doppler effect, giving a
velocity of recession, while magnitude $m$ gives a measure of distance
(how, and the units and order of magnitude of the distances, are
discussed below in Section \ref{sec:distances}). It was the distance
scale in Hubble's law that was in error.  In the 1950s, and
subsequently, that error was corrected, initially by a factor about 2.5
but now by a cumulative factor of order 7. Thus from 1952 onwards the
Friedman-\Lemaitre-Robertson-Walker (FLRW) expanding universes became
acceptable models for the real Universe though it was not until the
late 1960s that they became clearly the dominant
paradigm. \possessivecite{Nor65} philosophical and
\possessivecite{Ell89} historical account, which includes an
annotated bibliography, give more detailed background up to about
1960, and many more references to original papers and secondary
sources than there is space for here: see also \citeasnoun{Bon60}.

The incorrect distance scale was of less importance than the
revolution in humanity's picture of the Universe that the inferred
expansion implied.  It was obvious the heavens were not static, since
solar system bodies described forms of periodic motion, and it was
also known they did change, as shown e.g.\ by the Chinese observations
of the supernova that formed the Crab nebula, and by the comet of 1577
for which Tycho measured a parallax, demonstrating that it was moving
through the zone of the planets. Nevertheless, from Aristotle onwards,
much philosophical and religious thought considered the ``fixed
stars'' to live up to their name, so models of the large-scale
universe before Hubble's work were generally static: such a picture
was also supported, though not unambiguously, by the available
observations.

Despite the problem of the discordant ages, during the half-century
after Hubble's paper, expansion became fundamental to understanding the
nature and evolution of the matter in the Universe, in particular: the
formation of the chemical elements, from the combined effect of Big
Bang and stellar nucleosyntheses; the resulting inference of the
existence of at most 5, and probably 3, types of neutrino; and the
prediction of the Cosmic Microwave Background (CMB).

By 1980 there was therefore a well-established standard model, or
rather a set of models, FLRW universes containing pressureless matter
(``dust'') and radiation, which agreed with all the principal features
of the observed Universe as then known -- except one. The most obvious
fact about the Universe is that its density is not uniform -- it is
lumpy -- and within the models of the time this was explicable
only as coming from (rather unnatural) primordial irregularities.

Inflation theory, our current best explanation for the lumpiness, was
introduced in the 1980s. The generation of the necessary fluctuations
during a period of inflation, when the universe is driven by an unknown
field named the ``inflaton'' and is expanding very
rapidly, is now part of the cosmological standard model, the
``concordance model''. It involves quantum fields in the early
universe which produce a fluctuation spectrum, at the end of the
inflationary period, that matches the initial conditions required for a
subsequent classical evolution which gives the observed structures and
phenomena.  The theory of inflation is discussed in the companion
Milestone review, \citeasnoun{Dur15}.

The concordance model gives predictions for the spectrum of variations
in the CMB and their evolution. These predictions
depend on the behaviour of quantum and classical fields in an
expanding universe, and the evolution of perturbations during
expansion. The link between the spectrum of fluctuations at the end of
inflation and the present-day density variations is provided by the
(relativistic) theory of classical perturbations of the expanding FLRW
models. The resulting explanation of the density variations thus
represented an additional success for Einstein and Hubble. 

There was still a surprise to come. In the late 1990s two groups
announced, on the basis of the magnitude-redshift relation for
supernovae of type 1a (i.e.\ still using the same principles as
Hubble's study) that the Universe's expansion was accelerating. This
evidence, coupled in particular with the CMB observations and the
``baryon acoustic oscillations'' (BAO) found in galaxy redshift
surveys, led to our current picture in which the total energy density
of the Universe is close to the critical density, the boundary between
ever-expanding and contracting models, and made up of under 5\%
visible matter, about 25\% ``dark matter'' and about 70\% ``dark
energy''.

Additional evidence is, or is expected to be, available from many
types of observation such as gravitational lensing studies (using
another consequence of Einstein's theory: see \citeasnoun{Wil15}),
gravitational wave detection, observations of individual galaxies, and
terrestrial dark matter experiments, as well as from refinements of
the CMB, BAO and SN1a data.

It is interesting that although the results from the SN1a measurements
were unexpected, they could be regarded as theoretically predicted in
that it had been shown to be necessary to add a substantial $\Lambda$ to the
perturbed FLRW models to get agreement with observation
\cite{EfsSutMad90,OstSte95}.
 
There are still big open questions about the expanding universe, the
most obvious being the natures of the inflaton (the cause of
inflation), of dark matter and of dark energy.

This review will discuss the above points and aim to provide clear
indications of the fundamental importance of both relativity and the
Universe's expansion to our understanding of cosmology.

\section{Hubble's results}

\subsection{Observations preceding Hubble's}

For metaphysical reasons many people have had a strong bias towards a
static and unchanging Universe, albeit one including growth and/or change
in individual lives, in the motion of Solar System bodies, and so
on. Such views lay behind the development of the Steady State theory
\cite{Bon60}. However, it is important to realise that there were also
observational reasons for astronomers to favour a static universe.

The readily visible stars in the Milky Way, our Galaxy, occupy a
rather irregular but roughly coin-shaped region of what we now
understand to be the disk of a spiral galaxy. In Herschel's universe,
from 1785, these stars were taken to be the whole matter content of
the Universe, beyond which there was only empty space; and the Sun was
at the Universe's centre. We now know that this incomplete picture
arose because stars further away in the spiral arms of the galaxy, and
its bulge, are obscured from observation in visible wavebands due to
gas and dust. (The modern spectacular observations
\cite{EckGen96,GheSalWei08} supporting the presence of a supermassive
black hole at the centre of the Galaxy are made in the infrared.) The
extensive work summarized in \citeasnoun{Kap22} led to an ellipsoidal
model 3 kpc thick and of radius 15 kpc, with the Sun near the centre
(1 parsec, a pc, is $3.1 \times 10^{13}$ km $= 3.26$ light-years;
see section \ref{sec:distances}).

The alternative ``Island Universe'' concept, that the nebulae were
other star systems like the Galaxy, the viewpoint that Herschel had
hoped to confirm, was introduced by eighteenth century astronomers and
philosophers. It was supported by the nineteenth century resolution of
some nebulae into stars (see \citeasnoun{Nor65}), but was the
less-favoured option, by most astronomers, until the 1920s. The
arguments against concerned the relative sizes of the Galaxy and the
nebulae, and nebular spectra (an argument in which very different
types of nebulae were conflated).

The two decades leading up to Hubble's announcement saw a great deal
of work on distant stars and nebulae, covering the discovery of the
shape and size of our own Galaxy and the first redshifts and distances
of extragalactic nebulae. The numbers of references in the papers that are
cited in the following summary show how much more  was going on.

%
\citeasnoun{Sli14} had noted that the
shapes of spectral lines from spiral nebulae\footnote{``Nebulae''
  means clouds, in Latin: the distant agglomerations of stars looked
  like clouds in the telescopes of the time.} implied that those
nebulae rotated.\ van Maanen, initially in measurements of M101
\cite{van16}, alleged, on the basis of attempts to measure proper
motions, that the rotation periods were of the order of 85,000
years. (For a full account of van Maanen's work, and references, see
\citeasnoun{Het72}.) This could only make sense if those nebulae were
within the Galaxy. A similar inference was drawn from the 1885
observation of a supernova in M31, misidentified as merely a nova (it
was later used to infer a distance $\approx 200$ kpc for M31 by
\citeasnoun{Lun19}; although this was a significant underestimate it
sufficed to show that M31 was outside the Galaxy). That inference
helped to delay the recognition of the true nature of the other
galaxies in the Universe.

These viewpoints began to change due to other observations made from
the 1910s onwards. In 1912 Slipher became the first to measure the
redshift of an extragalactic nebula \cite{Sli13}, and found that M31,
the Andromeda nebula, approaches us at about 300 km/s. (It can be
argued from this and his subsequent papers that Slipher deserves a
large part of Hubble's credit for finding the expansion \cite{Pea13}.)

\citeasnoun{Sha18} measured the distances to a number of globular clusters
of stars and showed that they fill a sphere centred at the Galactic centre,
giving the first indication of the true shape and scale of the Galay
and the Sun's position within it. Globular clusters are visible out of
the plane of the Galaxy, but the absorption still present led Shapley
to overestimate the distance of the Galactic centre by a factor
2 (his figure was 20 kpc).

The two views on the nature of the nebulae gave rise to a ``Great
Debate'' between Shapley and Curtis in 1920 (see \citeasnoun{Tri95}),
Shapley arguing that the Galaxy was the whole Universe, and Curtis
arguing that the nebulae were ``island universes'', on the grounds of
the redshifts, the occurrence of dark regions like the dust clouds in
the Galaxy, and the rates of novae (Trimble discusses 14 points of
argument in total). The argument continued during the 1920s, during
which the balance shifted in favour of the nebulae being
extragalactic.

\citeasnoun{OorAriRoj24} showed that there was a halo of stars
round the Galaxy occupying the same sphere as Shapley's globular
clusters. Further confirmation of the size of the Galaxy came from the
work of \citeasnoun{Lin27} and \citeasnoun{Oor27}, who, in a series of
papers, proposed and observationally verified the differential
rotation of the Galaxy and thus its scale and the motion and position
of the Sun within it. This led to an estimate of 10 kpc for our
distance from the Galactic centre\footnote{Accurate measures of this
  distance are still difficult. The current conventional value is 8.5
  kpc.}. Hence the Sun could no longer be considered the centre of the
Universe. (The work by \citename{Tru30} \citeyear{Tru30,Tru30a}
directly measuring obscuration, and thus showing how the apparent
discrepancy over the size of the Galaxy arose, came after Hubble's
announcement.)

Slipher was building up the catalogue of known red- and blue-shifts of
nebulae. By 1917 he had 25, only 4 of them blueshifts
\cite{Sli17} and \citeasnoun{Edd23} was able to use 41, including 5 blueshifts.

\citename{Hub25} \citeyear{Hub25a,Hub25} obtained distances to M31,
M33 and NGC 6822 calibrated by observing variable stars (following
some work by Duncan), which he identified as Cepheids. (Interestingly,
Hubble obtained a less good estimate of the distance to M31 than had
been obtained by Curtis using observations of novae: see
e.g.\ \citeasnoun{Ste11} for references to this and other early
estimates.) In the first of these papers he noted that such stars had
already been detected in 3 more galaxies. He estimated the magnitude
of M31 as -21.8, corresponding to a distance of 285 kpc. In
\citeasnoun{Hub26}, which described his classification of nebulae, a
somewhat controversial matter \cite[Chapter 8]{Chr95}, he gives
distances to 32 galaxies and shows a relation between their absolute
magnitudes and that of their brightest stars. The continuation of this
work culminated in his 1929 paper.

It has been argued in retrospect that we should have realised there
was expansion, because it could explain the fact that the sky is dark
at night even though in an infinite static universe every line of
sight should end on a star and so be bright (Olbers' paradox). This
argument has been addressed and dismissed by \citeasnoun[chapter
  12]{Har81}.  He points out (a) that foreground stars obscure
background ones, so only a finite number of stars could be seen and
(b) with any reasonable lifetimes, stars cannot provide enough energy:
a calculation shows that the ``paradox'' requires the most distant
contributors to the light to be at $10^{23}$ light years.

\subsection{The theoretical developments: FLRW models}\label{oldtheory}

 The field equations of the theory of general
relativity (GR) can be written as
\begin{equation}
 G_{ab} := R_{ab} - \half R g_{ab}=\kappa T_{ab} + \Lambda g_{ab}
\label{EFE}
\end{equation}
relating the ``Einstein tensor'' $G_{ab}$ of a pseudo-Riemannian
spacetime to its energy-momentum content $T_{ab}$. The conventions and
definitions used above are defined in the next few paragraphs. Here
$\kappa= 8\pi G/c^4$, where $G$ is the Newtonian constant of
gravitation and $c$ the speed of light, in order to agree with
Newtonian gravity in an appropriate limit, and $\Lambda$ is the
cosmological constant. Einstein's initial version  of GR \cite{Ein15}
did not include the cosmological constant. He added it in
\citeasnoun{Ein17} precisely in order to have a static model of the
Universe.

The spacetime has a metric $g_{ab}$ of signature $\pm 2$ (the sign
choice is conventional), defining the scalar product of two tangent
vectors ${\bf v}$ and ${\bf w}$ at a point $p$ to be
$g_{ab}(p)v^aw^b$. The vectors' components here are given, in terms of
some suitable choice of basis vectors, $\{{\bf e}_a\}~(a=
1,\,2,\,3,\,4)$, by ${\bf v}= v^a{\bf e}_a$ for a vector ${\bf v}$.
Under gravity alone, test particles move on the geodesics of this metric.

The formulae relating the metric, the connection $\Gamma ^a{}_{bc}$
and the Riemannian curvature, in coordinate components where ${\bf
  e}_a = \partial/\partial x^a$, are
\begin{eqnarray}
\Gamma ^a{}_{bc}&=&\half g^{ad}(g_{bd,c}+g_{dc,b}-g_{bc,d}),\label{connform}\\
R^a{}_{bcd}&=&\Gamma ^a{}_{bd,c}-\Gamma ^a{}_{bc,d} +\Gamma
^e{}_{bd}\Gamma ^a{}_{ec}-\Gamma ^e{}_{bc}\Gamma ^a{}_{ed},\label{curvform}
\end{eqnarray}
where $g^{ad}$ is the inverse of $g_{bc}$. Here the subscript ${},b$
denotes a partial derivative in the ${\bf e}_b$ direction, while
${};b$ will similarly denote a covariant derivative.

The energy-momentum tensor of the matter content, $T^{ab}$, is assumed
to obey ${T^{ab}}_{;b}=0$; this generalizes the usual conservation laws to the
curved spacetime. 

The FLRW models are based on the Robertson-Walker metric, which can be
written as
\begin{equation}
\dx{s^2}=a^2(t)[\dx{r^2}+ \Sigma^2(r,K) (\dx{\vartheta^2}+\sin^2\vartheta\,
\dx{\varphi^2})]-\dx{t^2} ,\label{RWmetric}
\end{equation}
where $K$ can be normalized (by re-scaling as needed) to
$1,\,0~\mbox{\rm or}~ -1$. $K$ characterizes the three possible
curvatures of the hypersurface $t = \constant$ and $ \Sigma(r,K)=\sin
r,\,r~\mbox{\rm or}~\sinh r$ respectively.  The notation $a$ here
was used by \citeasnoun{LanLif41} and \citeasnoun{Lif46} and followed by
\citeasnoun{KraSteMac80}. It has become generally adopted. Other older
literature used instead $R$ for radius, $S$ for scale factor, or
$\ell$ for length scale.

This form is spherically symmetric about any point, and is homogeneous
on each $t = \constant$ hypersurface. \citename{Rob35}
\citeyear{Rob35,Rob36} and \citeasnoun{Wal35} independently showed
that any spacetime spherically symmetric about every point had to be
spatially homogeneous and admit this form of the metric. That is a
purely geometric result, and thus the form \eqref{RWmetric} is
commonly used in modified gravity theories as well as in GR.

The RW part of the FLRW name is thus slightly anachronistic in a
discussion of the development of the GR models before 1929. However,
by 1929 quite a number of solutions of \eqref{EFE} using
\eqref{RWmetric} were known, the most important contributions to the
expanding models being those of Friedman and {\Lemaitre} discussed
below: hence the name we now use. (Those exact solutions now known are
summarized in Chapter 14 of \citeasnoun{SteKraMac03} and references therein.)

For the metric \eqref{RWmetric}, the Einstein equations
\eqref{EFE} necessarily imply that the energy momentum has the perfect
fluid form\footnote{As stated, this form assumes the units are chosen
  so that $c=1$: for normal units one must replace $p$ by $p/c^2$.} 
\begin{equation}\label{Fluidform}
T_{ab} = \mu u_a u_b +p h_{ab},\quad h_{ab} := g_{ab}+u_au_b,
\end{equation}
where $u^a$ is a unit four-velocity; in \eqref{RWmetric} $u^a$ is
orthogonal to $t = \constant$. Assuming $\dot{a}\neq 0$, the equations
\eqref{EFE} reduce to
\begin{eqnarray}
3\dot{a}^2&=&\kappa\mu a^2+\Lambda a^2-3K,     \label{eq:11.1}\\
\dot{\mu}&+&3(\mu+p)\dot{a}/a=0.    \label{eq:11.2}
\end{eqnarray}
In honour of Hubble, $\dot{a}/a$ is denoted $H$.

\citeasnoun{Ein17} gave the Einstein static universe in which $K=1$
and $\dot{a}=0$. 
When $\dot{a}= 0$, $\mu$ is constant and one has to add the equation 
\[ \kappa(\mu +3 p) = 2\Lambda \]
to \eqref{eq:11.1}-\eqref{eq:11.2}.  It should be noted that
Einstein's static solution was a radical departure both from
observation and the two usual models of the day, Herschel's
and Island Universes, in that it assumed the Universe was
uniform and isotropic in space. It also introduced the description of
the matter content as a fluid, widely used since.

The other solutions with $\dot{a} = 0$ are forms of the empty spaces of
constant curvature -- flat space, de Sitter space ($\Lambda > 0$) and
anti-de Sitter space ($\Lambda <0$).
It was soon after Einstein's paper that
\citename{de17a} \citeyear{de17a,de17} found his eponymous metric,
which he gave in several sets of coordinates including
(in amended notation and units and with the opposite sign convention)
\begin{equation}\label{deSit}
\dx s^2 = -\cos^2 (r/R)\dx t^2 + \dx r^2 +R^2\sin^2 (r/R)(\dx \theta^2+\sin^2 \theta \dx
\phi^2).
\end{equation}
This solution has $\mu=0=p$, and $\Lambda = 3/R^2$. Note that de
Sitter did not give this in the form \eqref{RWmetric}, i.e.\ not in
coordinates referred to an expanding congruence. If that is done the
metric can be written as
\begin{equation}\label{deSit2}
 \dx s^2 = -\dx t^2 +{\mbox {\rm e}}^{Ht}
  (\dx \psi^2 +\cos^2 \psi[\dx \theta^2+\sin^2 \theta \dx \phi^2]),
\end{equation}
a form first found by \Lemaitre\ and by Robertson.

de Sitter was aware that there are redshift effects in this metric.
The exact form of those effects depends on which congruences of
emitters and observers are being considered, and some confusion arose
in early years because authors were not always careful to distinguish
between the possibilities. Underlying the ambiguity is the fact that
the de Sitter metric is four-dimensionally homogeneous and therefore
there is no naturally preferred set of worldlines or observers.

The principal choices were (a) observers on worldlines static in the
metric \eqref{deSit}, requiring some non-gravitational force to remain
in their positions, (b) freely-falling worldlines viewed by observers
static in \eqref{deSit}, and (c) the expanding congruence used in
\eqref{deSit2}. For (a) there is just gravitational redshift, as for
static observers in the Schwarzschild metric, for (b) the Doppler
shift of the emitters relative to the observers has to be added, and
for (c) the redshifts come just from the expansion. The three
approaches lead to different magnitude-redshift relations. \Lemaitre,
Weyl and others noted there would be a linear velocity-distance
relation in case (c).

\citeasnoun{de17a} called the Einstein static solution ``system
A'', and \eqref{deSit} ``system B'', names that remained in common use until
after 1929, and noted three redshifts (from Slipher and others), using
them to infer distances assuming the galaxies were static in system
B. Thus de Sitter's solution prompted the first analyses of the
redshift data, by several authors, in terms of a theoretical
model. For example, \citeasnoun{Edd23} used 36 redshifts and 5
blueshifts, mainly obtained from and by Slipher, some of them
otherwise unpublished at the time, while \citeasnoun{Lun24} assumed
that galaxies are standard objects, and deduced distances in units of
the M31 distance. Lundmark then showed a degree of correlation between these
distances and Slipher's redshifts, but with large scatter, and did not
interpret the results as showing an expanding universe. Although there
were no reliable distances, some authors noted a linearity between
velocity and distance.

Weyl's contribution is particularly notable in that he obtained a
general formula for redshift in any model and, in the 1923 fifth edition of
his book\footnote{According to \citeasnoun{Ehl09}, this edition has not been
  translated.} argued for a non-stationary model, considered an
expanding region within de Sitter space, and proposed  a distance
scale corresponding to an $H$ of 103 km/s/Mpc \cite{Ehl09}.

In 1922 and 1924, Friedman\footnote{Here I use the transliteration on
  his 1922 paper, which he used in later life. (I am grateful to
  Michael Heller for this information.) It is the more correct English
  transliteration from the original Russian. The commonly-used
  version, Friedmann, a German transliteration, appears on the 1924
  paper.} published his expanding universe models with positive
\cite{Fri22} and negative \cite{Fri24} spatial curvature\footnote{Both
  these papers have been translated and reprinted as
  \citeasnoun{Fri99}.}. \eqref{eq:11.1} is thus called the Friedman
equation. The matter content of these models was dust (so $p=0$),
formed by the energy-momentum of the ``gas'' of galaxies, taken to be
pressureless since collisions are infrequent.  (When discussing FLRW
models the dust constituent of the energy-momentum content is
sometimes just called matter.) It is a curiosity of history that the
zero curvature counterpart of the Friedman models, the Einstein-de
Sitter model, was found only in \citeasnoun{Einde32}, although
\citeasnoun{Rob29} had written down the general equations for all
three $K$ values.

\citeasnoun{Lem27} discussed the more general FLRW case with both dust
and ``radiation'' (a fluid obeying $p=\mu/3$, which arises naturally
from averaging over an isotropic distribution of
particles moving at the speed $c$, e.g.\ photons), as well as
$\Lambda$. It is this paper, together with Friedman's two, which
justify the FL part of the FLRW name. {\Lemaitre} set out to find a
model with nonzero matter content and a set of expanding
worldlines. He chose $K=1$, in order to have finite spatial extent,
and assumed the elliptic topology\footnote{This topology choice makes no
  difference to the dynamics.}. 

{\Lemaitre} then found the dynamic solution, later entitled the
Eddington-{\Lemaitre} solution, which tends to an Einstein static
universe as $t\rightarrow -\infty$ and to a de Sitter universe as
$t\rightarrow \infty$ (this is presented graphically in Figure
\ref{FLRWdyn}). Luminet's commentary on the reprint of this work (see
\citeasnoun{Lem13}) describes the evidence that {\Lemaitre} actually
calculated the behaviours of all the $K=1$ models, although he was
apparently unaware of Friedman's earlier work until 1929 when
Einstein told him about it.

{\Lemaitre} also derived the redshift formula for light observed by an
astronomer A from a source G in an FLRW (or just RW) model, where A
and G are assumed to be at constant spatial coordinates in
\eqref{RWmetric}: 
\begin{equation}\label{flrwz}
1+z = a_A/a_G.
\end{equation}
Apart from Weyl's book (see above), this was the first time redshift
had been related to an expansion of the universe rather than an
(apparent or real) motion of galaxies within a static spacetime. {\Lemaitre}
then related this formula to the known astronomical data (see the next
section).

{\Lemaitre}'s $\mu$ incorporated conservation of a total mass
$M$. \eqref{eq:11.2} is the remaining non-trivial Bianchi identity for
the metric \eqref{RWmetric}, and governs the evolution of the matter
content. Even today its major constituent is assumed to be dust,
representing the visible galaxies and invisible cold dark matter,
CDM\footnote{This name amuses British scientists because it was the
  abbreviation for the most popular UK chocolate brand.}. Here cold
means the matter's constituents have small kinetic energies compared
with rest mass, and thus exert only negligible pressure. The matter
content also includes the CMB, but this has a much smaller density now
than the dust. In a Big Bang model the dust and radiation had equal
densities at $t_{eq}$ after the bang: $t_{eq}$ is of the order of
$10^4$ yr.

In an FLRW model, dust has a density $\mu_d =
M_d/a^3$, where $M_d$ is a constant, and ``radiation'' similarly has
$\mu_r=M_r/a^4$. {\Lemaitre} thus had (in this notation) $\mu =
M_d/a^3 +M_r/a^4$, $p=M_r/3a^4$, but noted that $M_r$ could be
neglected when considering the application to astronomy.

We define the density parameter $\Omega_m := \kappa \mu/3H^2$, and 
similarly $\Omega_d$, $\Omega_r$, $\Omega_\Lambda = \Lambda/3H^2$ and
$\Omega_K = -K/a^2H^2$.
Note that necessarily (from \eqref{eq:11.1}) $\Omega_{total} =
\Omega_m+\Omega_\Lambda +\Omega_K = 1$.  At the end of inflation,
conversion of the energy-momentum of the inflaton
into radiation is assumed, giving $\Omega_r \simeq 1$,
with only very small contributions from $\Lambda$ and $K$. Because of
their dependences on $a$, $\Omega_K/(\Omega_m+\Omega_\Lambda)$ will
remain negligible during expansion unless $\Lambda =0$.  (A similar
remark applies for the forms of dark energy other than $\Lambda$ that
have been proposed.) That $\Omega_K/(\Omega_m+\Omega_\Lambda)$ is small
can thus be regarded as a testable prediction of inflation. Inflation
currently does not predict the present-day ratio of dark energy to
dark and luminous matter.

The deceleration parameter $q$ is defined by $q := - \ddot{a}/aH^2$, and obeys
\begin{equation}\label{qOmega}
 q = \half (\Omega_m+ \kappa pH^{-2}) -\Omega_\Lambda.
\end{equation}
Note that the sign adopted in the definition of $q$ was related to the fact
that if $\Lambda=0$ a positive $q$ was to be expected.

Many of the known solutions generalize the solutions for ``dust'' and
``radiation'' or a combination thereof, often assuming one or more
constituents with a barotropic equation of state $p=p(\mu)$, which is
frequently chosen to be of linear form, i.e.\ $p=w\mu$ where $w$ is a
constant. That form for the overall $\mu$ leads to
\begin{equation}
\label{qvalue}
q = \half \Omega_m(1+ 3w) -\Omega_\Lambda. $$
\end{equation}
Note that $\Lambda$ is equivalent to such a barotropic fluid with
$w=-1$, and that if $\Lambda=0$, $w=-1/3$ is the critical value
separating accelerating from decelerating universes.

As \citeasnoun{Hel74} pointed out, we still lack any detailed modeling
showing that the fluid approximation first used by Einstein is valid
throughout the different phases of the Universe's evolution. In
particular, while in the early universe there are very large numbers
of particles in small volumes, so the averaging usually implied by a
fluid approximation (see e.g.\ \citeasnoun{Bat67}, section 1.2) should
be valid, it is less than clear that this can be smoothly carried over
to a present day ``gas'' of galaxies, where averaging is over only
small numbers of particles. It may be that the unknown nature of the
cold dark matter  now inferred to be present throughout the Universe is such
that it resolves this issue.

More recently (minimally coupled) scalar field solutions with a potential
$V(\phi)$ have been widely studied: here the field $\phi$ obeys a field equation
\[ {\phi_{;a}}^a = \frac{\partial V(\phi)}{\partial \phi }\]
and in FLRW models gives an energy-momentum with
\[ \mu = \half \dot{\phi}^2 + V(\phi),\quad p =  \half \dot{\phi}^2 -
V(\phi)\] where the dot denotes $\partial/\partial t$.
Such fields have in particular been used to model the inflaton, the
dark matter, and the dark energy, and they may arise as effective
fields in (e.g.) considerations of averaged inhomogeneities
\cite{BucLarAli06}.

The perfect fluid form required by \eqref{RWmetric} can also be
constructed from two or more forms of matter not individually having
that form of energy momentum (see e.g.\ \citeasnoun{ColTup83}).

\subsection{Hubble's 1929 paper}\label{Hubble29}

As set out in the previous two sections, work on both the theoretical and
observational bases for expanding models of the universe had
gathered pace in the 1920s, and begun to interact, but Hubble's 1929
paper was undoubtedly the turning point.

\setcounter{footnote}{1}%
In {\Lemaitre}'s 1927 paper he, remarkably, used 42 galaxies, with
redshifts obtained from Str\"omberg and apparent magnitudes from
Hubble, to estimate the expansion rate. To do so he assumed, in the
same manner as \citeasnoun{Lun24}, that each galaxy has an absolute
magnitude equal to the mean of those whose absolute magnitudes had
been measured by Hubble (an assumption which inevitably increases
scatter) and found the relation we call Hubble's law. While this might
suggest renaming it {\Lemaitre}'s law, the results crucially depended
on Hubble's work (and Slipher's), and only in Hubble's work were
good\footnote{Up to overall scale.}
individual galactic distances used. Perhaps the right assignment of
credit is to Hubble for the observational facts and {\Lemaitre} for
the interpretation.

\begin{figure}
\begin{center}
\includegraphics{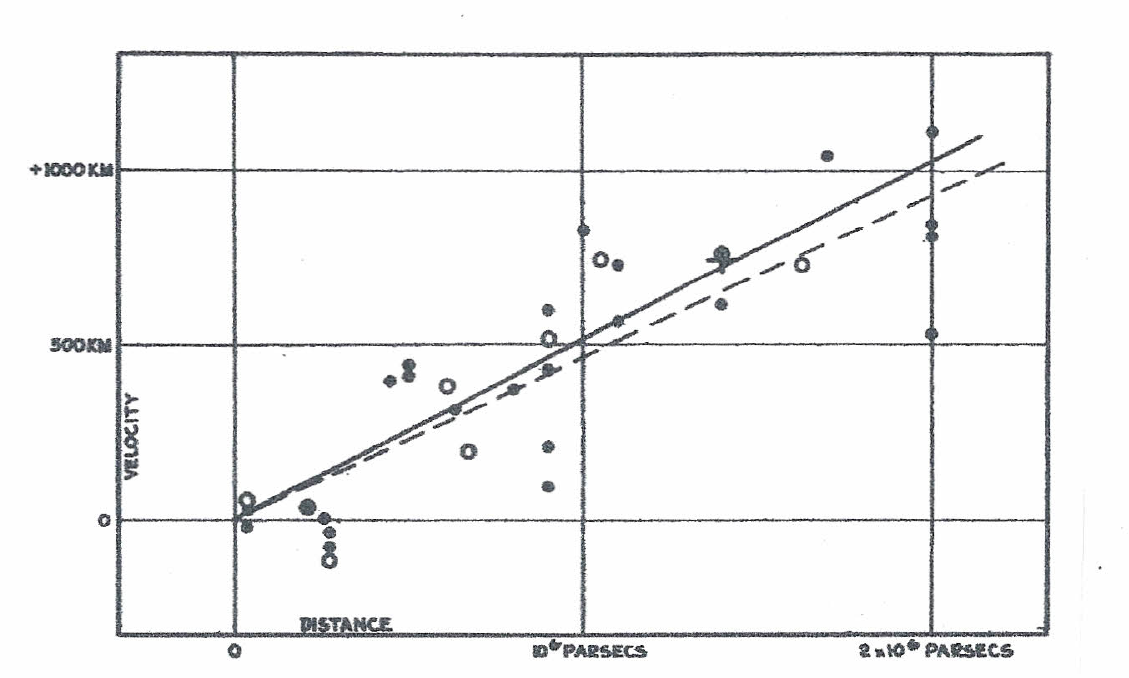}
\end{center}
\caption{[Hubble's original caption]\\
Velocity-Distance Relation among Extra-Galactic Nebulae.\\
Radial velocities, corrected for solar motion, are plotted against
distances estimated from involved stars and mean luminosities of
nebulae in a cluster. The black discs and full line represent the
solution for solar motion using the nebulae individually; the circles
and broken line represent the solution combining the nebulae into
groups; the cross represents the mean velocity corresponding to
the mean distance of 22 nebulae whose distances could not be estimated
individually. \copyright US National Academy of Sciences.\label{Hubblefig}}
\commentout{From PNAS web site:\\
Anyone may, without requesting permission, use original figures or
tables published in PNAS for noncommercial and educational use (i.e.,
in a review article, in a book that is not for sale) provided that the
original source and the applicable copyright notice are cited.}
\end{figure}

It was thus \possessivecite{Hub29} paper which gave a firm observational
basis for the linear relation between distance $d$
and recession velocity $v$ of galaxies, Hubble's law
\begin{equation}\label{Hubble}
v=Hd~.
\end{equation}
In the paper, Hubble plots the velocities and distances of 24 nebulae,
with speeds up to about 1000 km/s (i.e.\ a redshift around 0.03). To
obtain them he used the brightest star and Cepheid methods.  The most
distant four, which he identifies as members of the Virgo cluster,
were assigned distances about 2 Mpc. (Taking a modern value of about 70
km/s/Mpc for $H$, these galaxies are in fact at a distance of about 14
Mpc.) He also estimated an average distance for a further 22 nebulae,
using, as Lundmark and {\Lemaitre} had, the mean absolute magnitude of
the galaxies with measured distances and the measured apparent
magnitudes, and plotted that point; his plot is Figure \ref{Hubblefig}
(which he labels by velocity and distance rather than $z$ and $m$).

The coefficient $H$ in \eqref{Hubble} (which Hubble himself denoted
$K$) is known as the Hubble constant, although it varies with time in
observationally viable FLRW models: Hubble estimated it as 500
km/s/Mpc. (He actually found $H = 465 \pm 50$ km/s/Mpc from the 24
galaxies and $513 \pm 60$ km/s/Mpc by treating them in 9 groups.)
{\Lemaitre}'s analysis had been done with and without a weighting
intended to reduce the influence of the more distant galaxies on the
result, as the observations of those galaxies seemed less reliable,
and he found $H=575$ (without the weighting) and $H=625$ (with it), in
km/s/Mpc.

It is worth noting that, like Eddington and Lundmark, Hubble refers
only to redshifts in de Sitter space and not to the
Friedman-{\Lemaitre} expanding models. Other leading scientists
remained unaware of the expanding universe interpretation, or,
preferring a static picture, were initially unconvinced by it.  Hubble
himself is quoted by his biographer as saying in 1937 ``Well, perhaps
the nebulae are all receding in this peculiar manner. But the notion
is rather startling'' \cite[p. 201]{Chr95} and even in his Darwin
Lecture \cite{Hub53} talked of competing interpretations and said
the law should be regarded as ``an empirical relation between observed
data''.

The reasons why others failed to pick up on {\Lemaitre}'s discovery
are considered by Luminet in his excellent editorial note to the
recent reprint \cite{Lem13}; only in 1930, when {\Lemaitre} sent a
copy of his paper to Eddington, who then forwarded it to de Sitter and
Shapley, did the expanding model come to the fore. \citeasnoun{Edd30} also
discovered the instability of the Einstein static metric to small
changes of the parameters (whence his
name is attached to the Eddington-{\Lemaitre} model). This proof and
his advocacy of {\Lemaitre}'s work led to the wide acceptance of the
evolving models, although in \citeasnoun{Edd30} he noted that ``it is
possible that the recession of the spirals is not the expansion
theoretically predicted; it might be some local peculiarity...'', a
good point in view of the small sample then
available. \citeasnoun{Ein31} then renounced the cosmological constant
and considered a Big Bang model with recollapse \cite{ORaMcC14}.

In the English version of {\Lemaitre}'s 1927 paper, published in 1931,
the part on the magnitude-redshift relation is omitted. Research by
\citeasnoun{Liv11} found that this was not, as has sometimes been
claimed, a change made editorially, but was made by {\Lemaitre}
himself. Why that was done remains unclear. Luminet has discussed the
other changes made in the 1931 version.

\commentout{
Between 1928 and 1936 Humason made numerous additional distance
and (especially) velocity
 measures with the Mount Wilson 100 in telescope, out to as
far as $z=0.14$. in his 1936 paper he lists 100. He, Mayall and
Sandage gave a catalog in 1956
}

\section{Measuring the Hubble ``constant''}\label{Hmeasure}

\subsection{Measuring and interpreting redshifts}

Redshifts of galaxies can be rather accurately measured. This is done
by identifying patterns of lines in spectra which are characteristic
of emission or absorption by particular atoms or molecules, and
comparing the observed frequencies with those measured in the lab.

The method does of course assume that the emitted frequencies are not
affected by the evolution of the Universe, and some explanations have
abandoned that assumption. It has been examined in a number of works
over the years, the most recent being centred on discussions of possible
variations in the constants of nature, principally the dimensionless
fine structure constant $\alpha$. Variation in both time
\cite{MurWebFla03} and space \cite{KinWebMur12} has been claimed, the
latter showing a dipole which may be correlated with that of other
indicators \cite{MarPer13}. Such a variation can be described as a
variation of any of the dimensionful constants used to define
$\alpha$, e.g.\ the velocity of light $c$. The claimed, and still
controversial, observed variation is small enough that it does not
significantly affect measurement of the Hubble constant.

Treating light in the geometric optics approximation, one can show
that if $k^a$ is the vector tangent to a light ray from an emitter G
to an observer A, and objects have four velocities $u^a$, then the red-
(or blue-) shift observed by A is given by
\begin{equation}\label{redshift}
 1+z = (u^ak_a)_A/(u^ak_a)_G .
\end{equation}
If the $u^a$ are the velocities of timelike lines in the flat
(Minkowski) space of special relativity, then for purely radial
relative motion the redshift would imply a recession velocity $v$
given by
\begin{equation}\label{Doppler}
1+z = \sqrt{\frac{1+v/c}{1-v/c}},
\end{equation}
the Doppler effect. For $v$ small compared with $c$, $z \approx v/c$.

Since flat space is a good enough approximation to general
relativistic geometry out to a distance given by a square root of the
magnitude of the Riemannian curvature components, using the velocities
given by \eqref{Doppler} was a good enough approximation for Hubble and
remains a good one for considerably larger distances. (From
\eqref{mexpns} below, to a redshift $z>1$ with current $\Omega_m$ and
$\Omega_\Lambda$ values in \eqref{qvalue}.)

However, within general relativity the expanding universe models are
not flat. Since $k^a$ is geodesic, the change in the $(u^ak_a)$ of
\eqref{redshift} along a ray is due to the difference between
$(u^a)_G$ parallel-transported from G along the ray to A and
$(u^a)_A$. This has to be computed using the equations for geodesics
in the curved space and the assumed motions of the emitter and
observer, which are usually taken to be those of comoving or
``fundamental'' observers in an FLRW model, i.e.\ those at constant
spatial coordinates. For such observers this gives exactly the ratio
of the spatial distances between them at the times of emission and
reception, cf.\ \eqref{flrwz}; there has thus recently been some debate
about whether the redshift should be thought of as due to a Doppler
effect or not \cite{Kai14}. One aspect of the argument has been that
the calculation does not involve the Riemann tensor, only the
connection, and therefore should not be considered gravitational:
however, the Riemann tensor and its derivatives completely determine
local geometry, including the connection (see e.g.\ Theorem
  9.1 in \citeasnoun{SteKraMac03}).

One should note that galaxies and other sources with measured redshifts
will not be exactly following worldlines of fundamental observers in a
best-fit FLRW model. They will have ``peculiar motions'' relative to
that velocity. These tend to be small compared with the cosmic
expansion velocities, e.g.\ they are not more than a few thousand km/s
against the velocity of $\approx 2\times 10^5$ km/s at redshift 1
implied by \eqref{Doppler}. While thus not critical in deriving the
Hubble constant, measured peculiar motions are of great importance in
identifying large scale gravitating structures such as the Great
Attractor \cite{LynFabBur88}.

At the present time observed redshifts range up to 8.6 \cite{LehNesCub10}, and
candidates in the range of $z$ 8.5-12 have been identified \cite{EllMcLDun13}.

Spectral lines in visible light are not the only way to measure
redshift. One can also measure it from, for example, the 21cm line of
neutral hydrogen, common in both absorption and emission by
intergalactic gas clouds.

The origin of redshift in an expanding pseudo-Riemannian geometry is
clear, but there have been a number of controversies about this
FLRW interpretation. Apparent physical associations between objects of
different redshifts led to a prolonged debate \cite{FieArpBah73}. Other
authors thought there was a ``tired light'' contribution, or banding of
redshifts within clusters.

\subsection{Establishing the distance scale}\label{sec:distances}

Redshifts are relatively easily and well measured. Measuring the true
distances is much trickier: it is this which makes giving an accurate
absolute value for $H$ hard.

The full story of how astronomical distances are obtained involves a
great many astrophysical phenomena and techniques, and has
generated a series of subjects of debate over several decades. By
itself, it merited a very good book \cite{Row85}%
; and there have been subsequent developments, probably
including some of which I am unaware. So what follows is just a
short summary.

The principal idea, in measuring distance to other galaxies, is to
measure the flux received from a ``standard candle'' for which an
actual, intrinsic, luminosity $L$ (the total output, or the output at
given frequencies) is known, and then infer the distance by comparing
the observed flux and the intrinsic luminosity. Standard candles are
usually assumed to radiate isotropically.

There are several obvious problems with this procedure: sources do not
emit uniformly across different frequencies, so one wants to make the
measurements in such a way that the intrinsic luminosity and observed
flux refer to the same rest frequency; standard candles are not
strictly standard and one has to allow for the intrinsic variations
between members of a class, and possible resulting selection effects; and
to increase the distance range covered one has to use intrinsically
brighter standard candles, meaning one has to calibrate the intrinsic
luminosity of the new class from members of it which are clearly at
the same distance (e.g.\ because they are in the same galaxy) as
members of known fainter classes. The series of increasingly bright
classes form the ``distance ladder''.
 
For those new to this topic, the situation is complicated by
astronomers' adherence to magnitude and parsec, rather than SI,
units. Magnitudes $m$ arose in ancient astronomy. The Greeks
classed the brightest stars as being of first magnitude, the next
brightest as second magnitude, and so on: higher magnitude objects are thus
fainter. Because the eye responds essentially logarithmically to
received light, this means that the magnitude $m \propto -2.5
\log_{10} L$ for sources of intrinsic luminosities
$L$ at a fixed distance: the factor 2.5 was proposed by Pogson in 1856 so that 5
magnitudes corresponds to a factor 100 in luminosity, agreeing to a
good approximation with historic values for $m$.

To complicate matters, the zero of magnitude depends on wavelength,
and on the stars or other objects taken as calibration standards (see
e.g.\ \citeasnoun{Bes05}). The two main systems now in use are the
Vega (or Johnson) system and the AB system. The AB system is defined
so that zero magnitude is 3631 Jansky in every band, where 1 Jansky $=
10^{-26} W/m^2/Hz$.  The Vega system, originally defined so the star
Vega had zero magnitude in all bands, has been revised so that Vega is
now magnitude 0.02-0.03. The two systems are close in the visual V
band but, for example, in the near infra-red K band differ by 1.85
magnitudes. (As a historical note, a table produced by Hale Bradt in
1979 gave, for the wavebands U, B and V much used in the past for
photometric data, the values 1896, 4267 and 3836 Jansky in those
bands.)

Apparent magnitude $m$ is given by the measured flux, in $W m^{-2}$,
in the relevant waveband. Absolute magnitude $M$ is defined as the
apparent magnitude the source would have if at a distance of 10 pc. For
extended objects this has to be interpreted as the magnitude a point
source of equal luminosity would have at that distance. Then,
assuming an inverse square law for brightness, the ``luminosity
distance'' $D_L$ is given in pc by
$$ M-m = -2.5 \log_{10}(D_L/10)^2 \Rightarrow D_L/(10~{\rm pc}) =
10^{0.2(m-M)}.$$
The definition of luminosity distance implicitly assumes that there is
no redshift between G and A.  Redshift factors affect distance
measurements in two ways: they affect cross-sectional area
measurements of beams of light by relatively moving observers, and
produce spreading out of the spectrum.

If one defines the observer area
distance in terms of the cross-sectional area $\dx S_G$ of the
observed object G and the solid angle $\dx \Omega_A$ it subtends at A
by\footnote{The notation here differs slightly from that of
  \citeasnoun{EllMaaMac12}. Other authors denote $r_A$ by$D_A$ or $d_A$.}
$$ r_A^2:=\dx S_G/\dx \Omega_A,$$
then $D_L = (1+z)^2r_A$. Similarly one can define a distance $r_G$ from
the solid angle $\dx \Omega_G$ subtended at G by an area $\dx S_A$ at
A using $r_G^2:=\dx S_A/\dx \Omega_G$; then $D_L=(1+z)r_G$, $r_G=(1+z)r_A$.
When dealing with fluxes in specific intensity ranges (per Hz), one
also has to allow a $(1+z)$ factor relating the widths of the
wavebands. Past authors have not always been clear about what distance
measure they meant, and thus the various factors of $(1+z)$ were not
always treated correctly.

It should be noted that as a consequence of the relations between the
distance measures, a check on the value of $D_L$ is in principle
supplied by measuring $r_A$, assuming one knows $z$ and the intrinsic
cross-sectional area of a class of objects. The resulting analogue of
the $m-z$ relation is usually denoted the $\theta-z$ relation, where
$\theta$ is the measured angular diameter of the objects.

The first step in establishing the distance ladder is measurement of
trigonometric parallax, the same technique humans and other animals
with two eyes set apart use to infer distance. In astronomy one
observes the (small) change in the direction in the sky of an object,
usually a star, when the Earth is at opposite points of its orbit
round the Sun (i.e.\ 6 months apart). The parallax is then one half of
this, so it is $R_\oplus/D$ where $R_\oplus$ is the radius of the
Earth's orbit and $D$ is the distance. (For this purpose one can
assume Euclidean geometry.) One parsec (pc) is defined as the distance
at which an object in a direction perpendicular to the Earth's orbit
has a parallax of one arc-second. The known size of the Earth's orbit
($R_\oplus$ is roughly $1.5\times 10^8$ km) gives the SI units value
of this (to 1 d.p.) as $3.1 \times 10^{13}$ km: one parsec is 3.26
light-years. Distances within the Galaxy lie in a range from about 1.3
pc (the distance to Proxima Centauri, the nearest other star to the
Sun) to 20 kpc or so. M31 is at a distance of about 800 kpc. Distances
to the most distant galaxies measured by standard candle type methods
are of the order of Gpc. In modern galaxy surveys, Hubble's law is
inverted in order to infer (relative) distances from redshifts and so
derive separations and maps of structures.

Classical observations using parallax could lead to errors as much as
10\% at 30pc. A major target was and is the Hyades open cluster, a
group of stars used to calibrate methods used to make the next steps,
such as using stellar spectroscopy to identify stars which have common
properties and so can be standard candles, and the technique of main
sequence matching, described briefly below. The Hyades is the open
cluster nearest to us and is in the constellation Taurus: its stars
have been and are studied in great detail.

More recently, parallaxes have been measured using satellites such as
Hipparcos and the Hubble Space Telescope \cite{deHoode01,McABenHar11}:
the GAIA satellite which commenced observations in 2014 is expected
to enhance these considerably.  At
the distances accessible to direct parallax measurement, other methods
which are used as confirmations or checks include cluster parallaxes
(using apparent convergence of proper motions, i.e.\ tangential
velocities as seen from Earth) and statistical parallaxes of some
sample of stars. From these observations, the Hyades are now agreed
to lie at about 47pc.

Calibrating by Hyades stars, one can use stellar spectroscopy
for individual stars, and the position in the luminosity-temperature
diagram (the Hertzsprung-Russell diagram) of the zero-age main
sequence, the curve which newly formed stars settle to as they enter
the phase in which their energy mainly comes from fusing hydrogen to helium.

A next step in the ladder comes from the period-luminosity or
period-luminosity-colour relations for variable stars (the first of
these, for Cepheids, had been discovered by Henrietta Leavitt in
1908). RR Lyrae stars, W Virginis stars and Cepheids have been used,
with various difficulties of calibration. These indicators, together
with novae and supernovae, can be regarded as primary indicators,
enabling calibration of secondary indicators which can be used in
parallel to observe to even greater distances \cite{Row85}.

The supernovae considered as primary indicators are of two types
(these do not cover all supernovae). For both types, decay of unstable
radioactive isotopes, notably ${}^{56}$Ni, formed by the explosion is
an important, and for SN1a the only, source of the light.  Type II
supernovae, those with no hydrogen lines in the spectrum, are due to
core collapse of young massive stars, and are seen in spiral galaxies.
Theoretical models of surface brightness (in the simplest case,
assuming a black-body spectrum) and the observed luminosity can be
used to derive an angular size and this can then be combined with
Doppler redshift measurements to obtain the velocity of expansion and
thus the linear size and distance: this is called the Baade-Wesselink
method.

Type I supernovae have great homogeneity in the time variation of
their luminosity and colour. They are believed to be due to a white
dwarf exploding because accretion had raised its mass above the
Chandrasekhar mass limit: evidence for this has come from observations of
\citeasnoun{NugSulCen11}, but there is still uncertainty about the
origin of the accreted mass. Type Ia (the ones with a strong ionised
silicon absorption line) are the brightest: see section \ref{Modcos}
for the resulting $m-z$ data. Over 100 were known by the 1980s and new
ones are now being discovered at a rate greater than 2000 per annum.

Secondary indicators include ionised hydrogen (HII) regions, the
brightest globular clusters in a galaxy, the brightest stars in a
galaxy, the Tully-Fisher relation between the luminosity of spiral
galaxies and the width of the 21 cm emission of neutral hydrogen atoms
in those galaxies, and finally overall properties of galaxies
(colour-luminosity relations, luminosity classes, sizes, brightest
cluster galaxies).

Hubble based his distance measurements on observations of Cepheid
variable stars, calibrated by observations in the
Galaxy. Unfortunately, as I now discuss, the variable
stars he was observing in M31 and other galaxies were not Cepheids,
but the similar but less bright type of variable, the W
Virginis stars. He then used these measurements to calibrate the
``brightest star'' indicator, and so could find galactic distances
with both methods, but of course with incorrect overall scale. 

\subsection{Measured magnitude-redshift relations}

The first major revision of Hubble's results was brought about by the
commissioning of the Mount Palomar 200 inch telescope in 1950. Baade
used it to observe M31. Had Hubble's distance been correct, Baade
should have been able to see RR Lyrae stars, and he could not.  The
identification of two stellar populations\footnote{Populations I and
  II, to which are nowadays added the very low metallicity Population
  III stars.} arose in this work. The two populations are distinguished
by their metallicities\footnote{In astronomy, ``metallicity'' is the
  fraction of stellar matter not in the form of hydrogen or helium.}.

Baade realised that the two rather similar types of variable, W
Virginis stars and Cepheids, in the two populations had been conflated
by Hubble.  W Virginis stars (or type II Cepheids) are Population II
stars with brightest magnitudes around $-3.5$ whereas (type I)
Cepheids are Population I stars with brightest magnitudes about
$-5$. Disentangling the two led to revision of Hubble's distance scale
by a factor of about 2.5.

Because the principal uncertainty in magnitude redshift relations lies
in the distances, cosmologists often use a parameter $h= H/100$, where
$H$ is in km/s/Mpc. The Hubble age, $1/H$, is then approximately $10/h$
Gyr. Baade's $h$ was about 2. The next big revision arose from
Sandage's 1958 discovery that HII (ionized hydrogen) regions had been
misidentified as stars: he noticed that they looked too red to be
stars, as a result of the Balmer lines registering on red
plates. Sandage gave a value $h=0.75$; by 1975 he was using galaxies
out to redshifts above 0.3, i.e.\ ten times Hubble's most distant objects.

Baade and Sandage's discoveries not only ushered in a new phase in
cosmological science, as described in Section \ref{1959-80}, they also
provided the final step in establishing the ``Copernican Principle''
that humanity has no special position in the Universe. Copernicus and
Galileo had argued that the Earth was not the centre of the Solar
System, and Shapley, Lindblad and Oort had proved the Sun was not at the
centre of the Galaxy. The revised distance scales now showed the
Galaxy was not the uniquely largest in the Universe. (The recent
discovery of many planets orbiting stars other than the Sun, some
possibly habitable, provides further evidence against humanity having
a special position in the physical Universe: as yet the existence of
intelligent life on extrasolar planets remains conjecture rather than
confirmed fact.)

There was considerable debate about the correct value for $H$
after 1958, in which values between about 54 and 100 were obtained by
various sets of new observations and analyses. (Geoff Burbidge used to
show a log-log graph of the estimated age of the universe against the
date on which the estimate was made, which was linear starting from
Bishop Usher's estimate of 6000 years. He thus claimed that if it was
known when the next revision would occur one could predict its
value.) Nevertheless, the current best estimates essentially agree with
Sandage's 1958 figure.

It has been suggested that there are variations of $H$ with direction,
in particular a dipole variation due to the motion of the Galaxy with
respect to other galaxies. One difficulty in studying this lies in
relative distance scale calibrations for different directions.

The value of $H$ appears in a linear approximation for the $m-z$
relation at small $z$. Such a relation applies for any congruence of
worldlines of galaxies and observers with four-velocities $u^a$ in a
relativistic spacetime, and in this case $H = {u^a}_{;a}/3$. To study
the relation at greater $z$ one needs a more accurate
model. Observationally, it is natural to approximate the relation as a
series: the coefficient of the first nonlinear correction then gives
the value of $q$. The relevant expansion for sources of equal
intrinsic luminosity in an FLRW model reads
\begin{equation}\label{mexpns}
m = 5 \log z +1.086(1-q)z + O(z^2)+\mbox{\rm constant}.
\end{equation}
One can extend this series, and calculate series for other
measurements, not only for FLRW models but also for models with anisotropy and
inhomogeneity \cite{KriSac66,MacEll70}, but such approximations become
poor ones at the large redshifts now observed. It is more usual now to
compare data with numerically computed relations.

The numerous attempts up to the 1990s to obtain observations
sufficiently accurate to obtain $q$ from \eqref{mexpns} (as suggested
by \citeasnoun{Hub38}) ran into various experimental and theoretical
problems. These included the K-correction (relating fluxes from
different emitted frequencies), aperture correction (to make sure the
flux is solely and wholly from the intended object), absorption in the
Galaxy and the intergalactic medium, the effects of lumpiness arising
because the matter density within the observed beams of light differs
from the average density \cite{DyeRoe74}, selection effects (which
arise, e.g., from favouring the brightest in a standard candle class) and the
unknown evolution of the sources. Estimates varied from about 0.25 to
1.6.

The most prominent recent method for measuring the magnitude redshift
relation has used supernovae of type Ia. As described in section
\ref{Modcos} below, this led to the first good measurement of $q$, i.e.\ of
the time variation of $H$, giving a value which has since been shown
to be consistent with data from baryon acoustic oscillations and gamma
ray burst sources.

Yet more ways of obtaining and extending $m-z$ may become available
soon. For example, the use of reverbation mapping of AGNs and quasars,
which exploits the time delay between variations in the continuum and
line emission regions round AGNs and quasars and the theory of
accretion disks, could obtain sizes and hence distances, and gamma-ray
burst sources appear to extend the SN1a relation to much higher
redshifts.

It is perhaps disappointing that the correct value of $H$ is still
rather uncertain. Calibration of the SN1a data, still using Hubble's classic
method of observing Cepheid variables, leads to a value $73.8\pm 2.4$
\cite{RieMacCas11}, while the Planck data \cite{AdeAghArm14} gives
$67.3\pm 1.2$. This illustrates how hard it is to obtain precise
cosmological distances even now, 85 years after Hubble's work.

\section{Modeling the expanding universe, after 1929}

\subsection{Modeling Hubble's results}\label{1929-58}

There were many papers after 1929 exploring expanding universes,
though initially most avoided the singular ones, i.e.\ the Big Bang
cases. It was again {\Lemaitre} who took the lead in realizing the
significance of the singular models \cite{Lem31}. However, once the
concept of expansion had become generally accepted it was clear that
FLRW models agreeing with Hubble's data almost all expanded from a Big
Bang. Only the Eddington-{\Lemaitre} models asymptotic to the Einstein
static model as $t \rightarrow -\infty$ avoided a singularity, and
they were unstable.  There are models that ``coast'' close to the
Einstein static model for a period: these were invoked at one time to
explain an apparent clustering of quasar redshifts around 1.95 but no
such effect is now believed to exist.

\begin{figure}
\begin{center}
\includegraphics[scale=0.13]{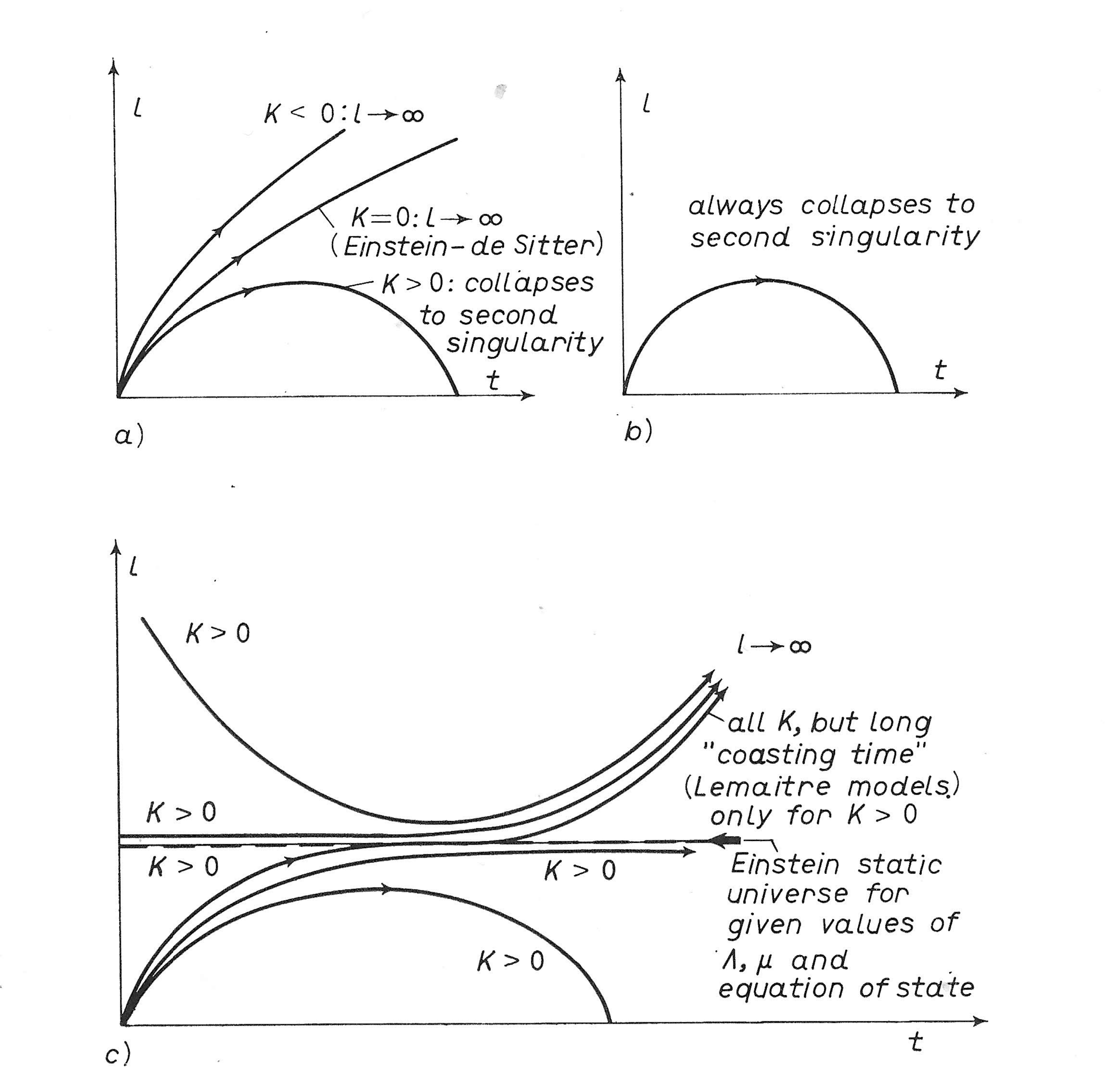}
\end{center}
\caption{Scale factor $a(t)$ for (a) $\Lambda =0$, (b) $\Lambda <0$,
  (c) $\Lambda >0$\label{FLRWdyn}.
From Figure 9.1 of \citeasnoun{EllMaaMac12}, itself adapted from Fig.\ 4 of
\citeasnoun{Ell71}.}
\end{figure}

Thus by the time of
\citename{Rob33}'s review \citeyear{Rob33}, the behaviour of all
the FLRW models was pretty well understood.
In summary, the generic behaviours possible are
illustrated by Fig.\ \ref{FLRWdyn}, assuming $\mu$ and $p$ are
positive in the case $\Lambda=0$.

At small $a$ the $\Lambda$ and $K$ terms in \eqref{eq:11.1} are
negligible. Then for $w=0$ (``dust'') we have $a \propto t^{2/3}$, and
for $w=1/3$ (``radiation''), $a \propto \sqrt{t}$; the corresponding
exact solutions are due to \citeasnoun{Einde32} and
\citeasnoun{Tol34}. For constant $w$ (other than -1) one obtains $a
\propto t^{2/3(1+w)}$. At large $a$, $\Lambda$, if non-zero, will give
the dominant term: if $\Lambda=0$, then, assuming that pressure
asymptotically vanishes as $a \rightarrow \infty$, the value of $K$
determines the far future behaviour (re-contraction if $K=1$,
expansion for ever if $K=-1$, approaching the Milne solution $a
\propto t$ which is flat Minkowski space in FLRW coordinates, and
convergence to the Einstein-de Sitter model if $K=0$).

As noted above, $\Omega_r$ is negligible for most of the life of the
Universe (for $t>t_{eq}$), although it is the dominant term between
the end of inflation and $t_{eq}$. On the other hand,
$\Omega_\Lambda/\Omega_d \propto a^3$ implies that a nonzero
$\Lambda$, or more generally dark energy in any of its proposed forms,
becomes dominant at late times in an FLRW expansion. Since
$\Omega_K/\Omega_d \propto a$, $K$ is dynamically unimportant at small
$a$, and for realistic FLRW models only becomes dynamically important
at large $a$ if the dark energy contribution decreases faster than
$\Omega_K$. Nonzero $K$ is the generic case, but the recent evidence
for nonzero $\Lambda$ has meant that the effort expended on
determining $K$ has lost part of its motivation, since a value $K \neq
0$ may not be critical in determining the future behaviour of the
expansion. That $\Omega_K/(\Omega_m+\Omega_\Lambda)$ is small is
consistent with inflation, as discussed above.

Taken together, these remarks, together with the considerations on
thermal history (see below and \citeasnoun{Dur15}), show why it is
usual to adopt a Tolman model in the early universe and a dust model
at late times (moderated, nowadays, by the inclusion of $\Lambda$ in
recent epochs).

The main reason why expanding models were not universally accepted
immediately after 1929 was that if \eqref{Hubble} were universal and
$H=500$ had remained constant as the Universe evolved, interpretation
as an expanding universe implied a ``Hubble age'' $1/H$ for the
Universe of 2 Gyr. With $\Lambda =0$ and positive $\mu$ and $p$ the
Hubble age is an upper bound for the age of an FLRW model. Hubble's
value for $1/H$ was
less than the age of the Earth, which was known to be about 4.5
Gyr. That discrepancy led to the consideration of various alternative
theories of gravity and cosmology, although most of the alternatives
also considered the universe to be expanding: for fuller accounts of
them see \citeasnoun{Nor65} and \citeasnoun{Bon60}.

Among the alternatives the most widely considered (at least in the UK)
was the Steady State theory first developed by \citeasnoun{BonGol48}
and \citeasnoun{Hoy48}. This used \eqref{deSit2} as the metric, so the
universe expands in this theory: it manages to remain the same at all
times due to continuous creation of new matter. Steady State theory
had nice simplifying features and produced definite predictions,
whereas Big Bang theory had some unknown parameters: accordingly
Steady State attracted a strong body of proponents.

Age problems involve not only the Earth's age.  Until the 1920s, the
only known source of the required energy for a star was the
gravitational potential energy of the star, but this ``contraction
hypothesis'' led to a maximum age around $2\times 10^7$ years, in
conflict with ``biological, geological, physical and astronomical
arguments'' (to quote Eddington). From about 1920 onwards (see
\citeasnoun{Edd26}) it was recognized that the power source was
nuclear fusion, for main sequence stars the fusion of hydrogen to
helium (another insight arising from relativity, in this case
Einstein's deduction within special relativity of the famous
$E=mc^2$). Estimates of stellar ages are subject to significant
uncertainties, due to the complexities of modelling stellar
evolution. The greatest currently estimated age of an observed star is
$14.5 \pm 0.8$ Gyr \cite{BonNelVan13} which is very close to current
estimates of the time back to the Big Bang.

Baade's distances led to a Hubble age of 5 Gyr, just about compatible
with the age of the Earth. The age of the Galaxy was then estimated to
be between 3 and 15 Gyr, on the basis of data from meteorites, stellar
evolution and dynamics, and observations of interstellar dust
\cite{Bon60}, so this too could just about be accommodated. These age
estimates became comfortable after Sandage's revision, which led to a
Hubble age of about 13 Gyr and removed that motivation for alternative
theories.

Despite the age problem, FLRW models continued to be investigated
between 1929 and 1952, if
perhaps more slowly than they might have been, and some aspects which
have since assumed ever-increasing significance were first explored. 

The idea of remnant radiation from the Big Bang arose in
\citeasnoun{Lem31}, but the resultant black-body spectrum and
decoupling were not discussed then and {\Lemaitre} thought cosmic rays
might be the remnant. Our current picture is of course that baryonic
matter and radiation are strongly coupled in the early universe,
that the relevant temperatures therefore evolve together, and that the
coupling ends at a temperature around 3000K (more accurately, at about
$z=1089$) although the radiation remains thermal (black-body). After
decoupling, the radiation propagates freely (or, possibly, interacts
with a reheated intergalactic medium). The decoupling happens rather
abruptly and thus can be characterized as happening at a specific
surface $t=t_{dec}$, the ``last scattering'' surface.

Tolman considered the thermodynamics of Big Bang models, including the
effect of expansion on black-body radiation temperature and of a gas
in equilibrium with it \cite{Tol34}. However, he did not infer the
presence of the CMB. That was predicted later, when the lack of
equilibrium, which is forced by expansion \cite{SteMacSci70}, was
brought into play to discuss element formation, in work which also led
to the theory of primordial nucleosynthesis.

Such nucleosynthesis was first considered by \citename{Gam48}, in a
series of papers from 1942 onwards. He recognized the need for high
temperatures to enable neutron capture, and the possibility of this
for a limited period in the expansion of radiation-dominated models
\cite{Gam46,Gam48}. Alpher, Herman and colleagues developed this
further, and carried out detailed nucleosynthesis calculations. They
also predicted the CMB at a temperature of 5K \cite{AlpHer49}. The
outcome of these calculations was that primordial nucleosynthesis
could produce the light elements but not the heavy ones
\cite{AlpFolHer53,Gam56}.

The mechanism to form light elements starts with neutron-proton merger to form
deuterium followed rapidly by a series of reactions to form
$He^4$. The deuterium is instantly dissociated by incident photons if
$T \geq 8\times 10^8$K. Neutrons decay to protons with a half-life
about 660s and after this has happened no more deuterium forms. So the
outcome depends on the neutron-proton ratio when $T=10^8$K, which gives
the initial condition for deuterium formation, and the time scale of
expansion, which causes collision rates to drop as density does. It was
\citeasnoun{Hay50} who pointed out that the initial ratio arose from
processes with high $T$ dependence which would give a thermal
equilibrium until their time scales became slow compared with
expansion. Taking that into account the helium to hydrogen ratios came
into good agreement with observation. \citeasnoun{WagFowHoy67} refined
the calculations and included elements up to Li${}^7$.

The theory of production of heavy elements in stars was developed by
\citeasnoun{BurBurFow57} in the context of the Steady State
theory. This work showed such synthesis could not produce the observed
deuterium. Stellar nucleosynthesis complements the primordial element
production in giving the observed species.

Another important topic opened up in this period was the theory of
perturbations of expanding models. Newtonian models were 
considered by various authors, notably \citeasnoun{Bon56}: \citeasnoun{Lif46} was the first to
obtain a growth law for perturbations in relativistic models and
introduced methods followed in later work. However, until 1980 there
were no plausible ways to generate the necessary perturbations. For
example, in giving a summary of what was then known,
\citeasnoun{Har73} noted that if thermal fluctuations provided the
seeds for galaxies, they had to be generated between $10^{-31}$ and
$10^{-29}$ s after the Big Bang. 

Yet another topic introduced in this era was the study of inhomogeneous
and anisotropic models (see chapters 17-19 of \citeasnoun{EllMaaMac12}
for a recent review of these). \citeasnoun{Lem33} introduced the
spherically symmetric dust models now called the
{\Lemaitre}-Tolman-Bondi (LTB) models, in order to model collapse of
overdense regions into galaxies. In the same paper he considered,
after a suggestion from Einstein, the spatially homogeneous anisotropic
Bianchi I models, in order to illustrate that the singularity at the
Big Bang of the FLRW models was not unique to that case.

\subsection{Modeling with improved distance scales}\label{1959-80}

In the 1960s two observational developments in particular, the counts
of radio sources and the detection of the CMB, gave the Big Bang view
the edge over Steady State (although Steady State enthusiasts
continued for some time to try to incorporate these observations
within their theory).

The cosmic radio source count data showed that the universe was not only
expanding but evolving. It did so by plotting source count numbers $N$
against radio flux $S$, so proving that the number of radio
sources in the past was higher than today by more than could be
accounted for by geometric effects in any reasonable model.  The data
from which this inference was initially drawn was inadequate, due
to confusion between sources etc.\ but by 1968 these systematic errors
had been corrected \cite{PooRyl68}; see Figure 3.
\begin{figure}
\begin{center}
\includegraphics[scale=50]{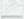}
\end{center}
\caption{The $N-S$ curve from \citeasnoun{PooRyl68}, Fig. 2, and its
  caption\label{pooley}} 
\end{figure}
Therefore the population of radio sources, and thus the
Universe itself, had to be evolving. Unfortunately, the intrinsic variations
between radio sources and the lack of detailed understanding of their
evolution makes it hard to gain other cosmological information from them.

The earliest paper to re-examine the CMB prediction was by
\citeasnoun{DorNov64} but, as described by \citeasnoun{Nov09}, its
significance remained unappreciated, as did data by Shmaonov and the
observations of the CN molecular lines. For discussion of the
discovery by Penzias and Wilson and subsequent observational and
theoretical work see \citeasnoun{Dur15}. The presence of the CMB confirmed the
evidence from element abundances, showing that the universe had
been through a hot dense phase, and had thus come from a Big Bang.

The generally accepted thermal history for FLRW models now started
with a
\commentout{
hadron era, before $10^{-5}$ s, followed by the lepton era at
$10^{-5}-10$s. The particles present in the lepton era were all
known. If massive, they were assumed to have high energy and so be
effectively massless. Adding up the contributions of the different
particle species led to $\mu = 9aT^4/2$ where $a$ is Stefan's constant
$k^4\pi^2/15c^3\hbar^3$.  This gave a}
Tolman radiation universe with
known ratio $a/\sqrt{t}$ (see \citeasnoun{Dur15}),%
\commentout{There was $e^+-e^-$ annihilation at $\approx
1$s, after which $\mu=11aT^4/4$. One could infer from the CMB
temperature that there were relic electron neutrinos (assumed
massless) with a present day temperature of 2.1K and relic
gravitational waves with $T=1.6$K.

Next came the plasma era, until $10^{12}$s, during which element
formation took place.}
followed by a ``matter era'' driven by dust but
with unknown $\Omega_d$.

Isotropy about our position in space was shown at that time by a
number of tests. Galaxy number counts were only known to be isotropic
to about 30\% (the uncertainty being due to the clustering), but the
radio source counts and X-ray background were isotropic to better than
5\% and the CMB was known from about 1969 onwards to be isotropic to better
than 0.2\%. The first anisotropy measured in the CMB, in 1969, was the dipole
due to the motion of the Sun relative to the frame in which the CMB is
most nearly isotropic (see \citeasnoun{Dur15} for details and references).

In contrast, then and now, obtaining clear evidence for spatial
homogeneity is difficult in that we do not receive data from objects
separated from us in space but not in time. In particular we cannot in
principle separate space and time evolution, since, apart from
``geological evidence'' in our local neighbourhood, our observations
are on our past null cone. So to infer homogeneity implies assumptions
about how the matter content behaves away from the null cone. It is
not even easy to test the assumption that there are spacelike surfaces
of approximate homogeneity extending to large distances. For example
the surface of last scattering might be a timelike cylinder
intersecting our past null cone in the sphere we actually observe
\cite{EllMaaNel78} (though the particular model where this was
considered is ruled out by other considerations).  Even when one takes
the more usual assumption that last scattering occurs on a spacelike
surface, one needs to be able to apply some theory of the evolution
objects undergo between the emission of the received radiation and
now.  These assumptions are hard to test.

Thus to improve tests of homogeneity we would need to have a much better
understanding of evolution of galaxies, radio sources and other
objects than we do. We are however able to make two types of tests: we
can check that there is no local evidence for inhomogeneity, for as far
back in time as we feel confident about our understanding of evolution,
and we can test for the presence of significant mass variations which
are asymmetrically arranged about us by looking for their effects on
the CMB. The tightest bounds of the latter type seem to be those of
\citeasnoun{ZibMos14}.

Thus in the 1960s and 1970s it was becoming ever clearer that the FLRW
models satisfied the observational requirements of expansion,
evolution, a hot dense phase, and apparent isotropy and spatial
homogeneity. They accounted for the evolution of the light elements,
while nucleosynthesis in stars accounted for the heavy ones, and they
naturally gave rise to the CMB. The value of $H$ was believed to be at
most 100, and therefore there was no serious age problem.

The models still had significant unknown or poorly known parameters:
$\Omega_d$, $K$, $q$ and $\Lambda$. They also did not predict the
baryon number $B$, the ratio of the number of photons per unit volume
(in the CMB) to the corresponding count of baryons, which is a measure
of entropy and affects the helium/hydrogen ratio and other light
element fractions, and so on.  There were attempts to obtain the FLRW
model parameters not only from the $m-z$ relation but also the
$\theta-z$ and $N-S$ relations described above, the $N-m$ relation for
galaxies (a test proposed by \citeasnoun{Hub36}), and the
$<V/V_{max}>$ test, where $V$ is the volume of a sphere of radius
given by the distance of a source and $V_{max}$ the maximum volume
within which it would be detectable: precise values remained elusive.

It became clear that there was dark matter in the universe
(see \citeasnoun{LonRee73} or \citeasnoun{ColEll97}).
The phrase ``missing matter'' came into use, but rather confusingly
was used both for the mass inferred to be in galaxies and clusters in
addition to their visible mass (which pointed to an $\Omega_m$ of about
0.3: see the references immediately above) and by those who believed
that the true $\Omega_m$ was 1, although there was no observational
evidence for such a large value (or for a non-zero $\Lambda$).

The period saw some important steps in understanding light propagation
in the Universe. \citeasnoun{SacWol67} worked out the effects on the
CMB of gravitational redshifts: the fully non-linear integrated
effects were discussed by \citeasnoun{ReeSci68}.
\citeasnoun{SunZel70} described the effect of inverse Compton
scattering in galaxies on CMB observation.  See
\citeasnoun{Dur15}. Several papers, notably those of
\citename{DyeRoe74}, considered light propagation in lumpy models.  

One aspect of FLRW models studied in greater detail was the impact of
the exact rate of expansion during element formation (which depends on
the particles present and on the anisotropy) on the resultant ratio of
atomic species.  \citeasnoun{SteSchGun77} showed that such
cosmological considerations limited the number of neutrino species, at
that time unknown but expected from the quantum theory of the standard
model of particle physics to be 3, to at most 5. A more recent
re-analysis \cite{CybFieOli05} gave $2.67 \leq N_\nu \leq 3.85$ at
the 68\% confidence level, while Planck's data gave $N_{eff}= 3.30 \pm
0.27$ \cite{AdeAghArm14}. Similarly, \citeasnoun{Bar76} used the
effect of anisotropy on the expansion rate to give a limit $10^{-7}$
on present-day anisotropic shear $\sigma$, possible because a
$\sigma^2$ term has to be added to the Friedman equation
\eqref{eq:11.1}.

Big Bang FLRW models have a singular origin.  The understanding of
singularities, defined as the presence of geodesics which could not be
indefinitely continued, developed considerably during the 1960s and
70s. Theorems proving their existence in relativity (see
\citeasnoun{HawEll73}, \citeasnoun{TipClaEll80} and the
Milestone ``The singularity theorems (1965)'') were complemented by
examples of possible behaviours and by the work reviewed in
\citeasnoun{BelKhaLif82} aimed at describing singularities' generic
form. As a result of the CMB discovery, it was possible to prove that
a relativistic cosmology that was expanding approximately like an FLRW
model had to have had a trapped surface in the past and therefore must
have had a singularity in the past \cite{EllHaw68}.

\section{Modern relativistic cosmology}\label{Modcos}

\subsection{Observations}

Three important sources of observational data, the anisotropies in the
CMB, the $m-z$ relation itself, and galaxy surveys, have undergone
major developments since the 1980s. Combining the results has given a
much stronger base for our modeling.  Moreover, other sources of data
have been and are being added.

The first to improve was the measurement of the fluctuations in the
CMB (beyond the level of the dipole due to the Earth's motion),
discussed more fully by \citeasnoun{Dur15}. This has
evolved from the first indications in the COBE data of the early 1990s
through to the very detailed recent results by Planck
(e.g. \citeasnoun{AdeAghArm14}) and a number of ground-based
telescopes. There are ongoing observations by the ever-increasing
number of ground-based instruments which will give us further fine
detail.

Drawing the implications from this data depends on calculating the
evolution of fluctuations from the input fluctuation spectrum predicted
to have been formed during inflation, within an FLRW model (see
section \ref{StanMod} and \citeasnoun{Dur15}). Comparing
such calculations with the observations puts constraints on the
expansion parameters $H$, $\Omega_m$, $\Omega_K$ and
$\Omega_\Lambda$. Note that these parameters give rise to accumulated
effects over a long period, rather than referring to measurements of
relatively nearby objects.

The most unexpected addition to previous data came from the Hubble
relation for supernovae of type 1a. It was unexpected because if
$\Lambda=0$, as was generally believed, then \eqref{qOmega} shows that
the expansion rate should be decelerating.  The result, first
announced in 1998, showed instead that $q<0$, implying $\Lambda>0$: it
won the lead scientists the 2011 Nobel prize.

The CMB observations show a first main peak in the power
spectrum, from which a length scale at the time of the last scattering
of the CMB can be inferred. Expansion will then fix the corresponding length
scale for a peak in the distribution of galaxy-galaxy separations at a
later time. This is the BAO measurement, made in surveys of
galaxies. Taking galaxies in a given band of redshift, around a value
$z_B$ say, allows one to estimate the evolution between the time of
last scattering and $z_B$ and thus the $\Omega$ values, assuming an
FLRW model.

The BAO peak is a less than 1\% deviation from a uniform
distribution. Nevertheless it can be measured with considerable, albeit
model-dependent \cite{RouBucOst15}, accuracy. The same peak has
recently been measured using the Ly$\alpha$ forest in the spectra of
quasars at redshifts in the range $2.1 \leq z \leq 3.5$
\cite{DelBauBus15}. Using this data and a value for the sound horizon
obtained from the Planck data leads to a value for $H$ at $z=2.34$ of
$222 \pm 7$, about 7\% ($2.5\sigma$) discrepant with a flat $\Lambda$CDM
cosmological model with the best-fit Planck parameters.

\begin{figure}
\begin{center}
\includegraphics[scale=0.4]{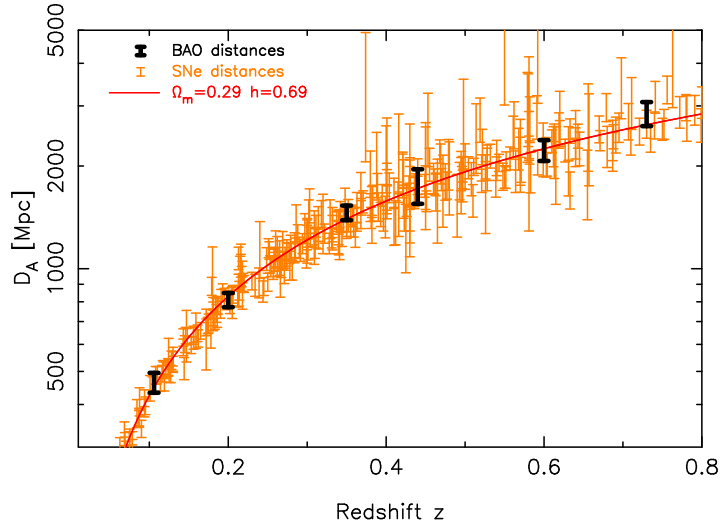}
\end{center}
\vspace*{-1cm}
\caption{[Caption from \citeasnoun{BlaKazBeu11}]
Comparison of the accuracy with which supernovae and baryon
  acoustic oscillations map out the cosmic distance scale at $z <
  0.8$.  For the purposes of this Figure, BAO measurements of $D_V(z)$
  have been converted into $D_A(z)$ assuming a Hubble parameter $H(z)$
  for a flat $\Lambda$CDM model with $\Omega_{\rm m} = 0.29$ and $h =
  0.69$, indicated by the solid line in the Figure, and SNe
  measurements of $D_L(z)$ have been plotted assuming $D_A(z) =
  D_L(z)/(1+z)^2$.\label{WiggleZ}}
\end{figure}

The accumulated SN1a data has been compared with the BAO measured by
the WiggleZ team
\cite{BlaKazBeu11} at several redshifts out to $z>0.7$ and agrees very
well; see Figure \ref{WiggleZ}.

\begin{figure}[h!]
\begin{center}
\includegraphics[height=3in,width=3in]{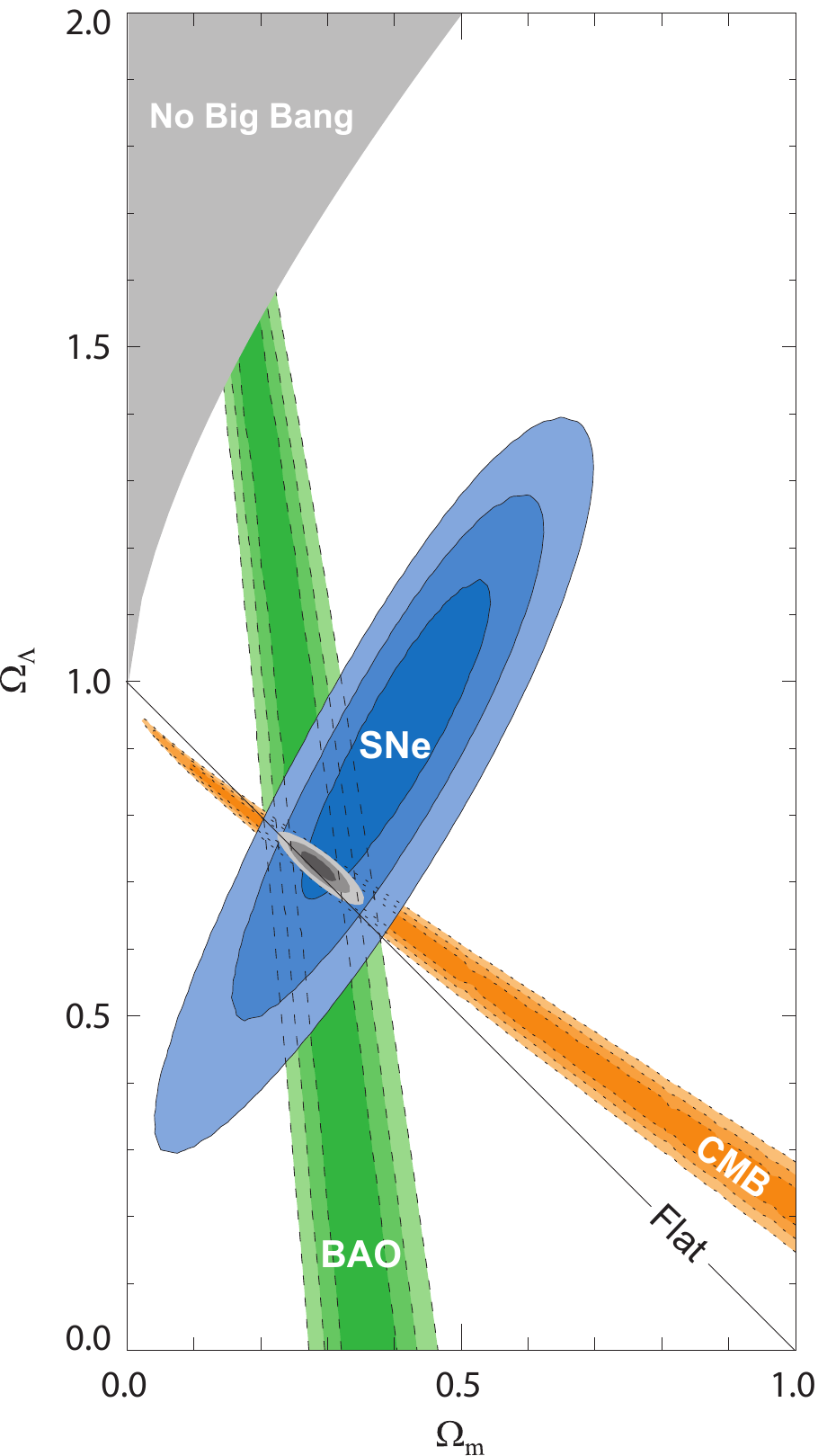}
\end{center}
\caption[Joint constraints on $\Omega_{{\rm m}0},\Omega_{\Lambda 0}$ from
SNIa, BAO and CMB]{\small{Constraints (68\%, 95\% and 99\% CL contours) in the
$(\Omega_{{\rm m}0},\Omega_{\Lambda 0})$ plane from
SNIa, BAO and CMB.
(From \citeasnoun{KowRubAld08}.)}} \label{omlm}
\end{figure}

Values for $\Omega_m$ and $\Omega_\Lambda$ can be obtained by
combining the SN1a, CMB and BAO data as shown in Figure \ref{omlm}.

Further important input comes from weak lensing. The images of distant
sources are distorted by the light-bending due to matter nearer to
us. This distortion may be strong, for example causing multiple images
of a single lensed galaxy, if the light passes very close to the
lensing object, but for many cases only distorts what would be a circle
into an ellipse: typically the ellipticity only changes by about 1\%.
We do not know the actual shape of the lensed object.
Nevertheless, statistically one can extract the density distribution
of the mass causing the lensing.

The first map of dark matter using weak lensing (with a nice
three-dimensional picture) was given by \citeasnoun{MasRhoEll07}, in a
1.6 square degree area. The more recent data, with a method
complementary to the statistical calculations
\cite{VanBenErb13}, shows detailed maps of the (dark and
visible) matter in regions several degrees across.  Several lensing
surveys are in progress to improve this data to the level where it may
also be able to give additional information on dark energy.
Lensing has also been detected in the CMB measurements \cite{AdeAkiAnt14}.

\subsection{The standard model and the development of structure}\label{StanMod}

The current standard model of cosmology is an expanding approximately
FLRW Big Bang model, with, taking the values given by the Planck data
\cite{AdeAghArm14}, $\Omega_m \approx 0.31$, of which about 0.05 is
baryonic matter and the rest dark matter, and $\Omega_\Lambda \approx
0.69$. It is thus known as the $\Lambda$CDM model. These $\Omega$
values do not require revision of our understanding of thermal
history, since $\Omega_\Lambda \ll \Omega_m$ for $z>1$.

In the early universe, the standard model undergoes a period of
inflation, with a rapid expansion driven by an inflaton, a field not
(so far) observed terrestrially: exact details are model
dependent. The crucial role of this early period is that it leads to a
spectrum of classical density perturbations that is almost flat. The
inflaton is typically modeled using a scalar field.  This can give
rise to a range of behaviours for $a(t)$ depending on $V(\phi)$.

The thermal history of this model is described in \citeasnoun{Dur15}
and will not be repeated here: it includes the nucleogenesis and other
processes on which expansion exerts a major influence as described above.

The development from the initial fluctuations to the last scattering surface
and later times is modelled by perturbations of the FLRW
background. This simple statement hides two problems, the choice of
the FLRW model to perturb (the ``fitting problem''), and the
``gauge'' issue which arises from the lack of an invariantly-defined
map between the real lumpy Universe $M$ and the fictitious smooth FLRW
universe $M'$.

The gauge issue is often described in coordinate terms, and then
results in changes of variables similar to those of gauge
transformations in field theory: hence the name. It is usual to
identify points $p \in M$ and $p' \in M'$ which have the same
coordinates, say $x^a=x'^a$. Making the same coordinate change in each
manifold does not change this identification. If one slightly alters
the map, i.e.\ the choice of $p'$ for a given $p$, this can be
described as mapping $x'^a$ to $x^a= x'^a+\delta x^a$, a change of
gauge. Since this is clearly not a physical change in the real
universe, one wants a description independent of the gauge choice.

\citeasnoun{SteWal74} showed that the only gauge invariant quantities
are those which in the background are zero, constant scalars, or sums
of multiples of the tensor $\delta^i_j$. This is a nuisance, and in
particular implies that there is no gauge-invariant way to define the
perturbation of density in an FLRW model, since density is a
time-dependent scalar.

The fitting problem of picking the best-fit FLRW model for the
comparison also involves gauge. The perturbed model could, for
example, have greater average density than the initial model and so
demand a different best fit. Fitting is also closely related to the
``averaging problem'' which is as follows. The equations
\eqref{connform}-\eqref{curvform} are nonlinear, and hence the
averaged curvature is not the curvature of the averaged metric or
averaged connection (because the average of a product, $<ab>$ say, is
not the product of the averages $<a><b>$). This implies that the
averaged density of the real universe may not be the same as that
implied by \eqref{eq:11.1} for the FLRW model whose average $H$ agrees
with that of the real universe. One can try to estimate the effect by
calculating the difference between the average of the curvature and
the curvature of the averaged metric, the ``backreaction''.

Unfortunately we do not yet have a generally-agreed way to do
this. The technical difficulties include the need to compare tensorial
quantities at separated points, and the necessity of defining
averaging volumes in a non-covariant way (since Lorentz
transformations map a point at a finite distance to points on a
hyperbola stretching to infinity). The various attempts are reviewed by
\citeasnoun[chapter 8]{Kra97} and \citeasnoun[chapter
  16]{EllMaaMac12}, q.v.

A further difficulty is that a best-fit RW metric (i.e. of form
(\ref{RWmetric})) might not show the time evolution of an FLRW model with
reasonable matter content.

One approach much explored, because only averaging of
scalars is required and this is well-defined, is that of Buchert: it
involves averaging for fundamental observers  and the non-commutation
of the time derivative and averaging operators, leading to modified
Friedman and Raychaudhuri equations (see
e.g.\ \citeasnoun{Buc08}). Because tensorial quantities are not
averaged, this approach has to be completed by adding physical closure
conditions.

There are two main approaches to calculating the behaviour of
perturbations, the metric-based method of \citeasnoun{Bar80} which develops
the ideas of \citeasnoun{Lif46} into a gauge-invariant treatment, and the
covariant approach of \citeasnoun{EllBru89} which developed ideas of
\citeasnoun{Haw66}. There are quite a large number of papers where
these are expounded and applied: these are surveyed and summarized in
\citeasnoun[chapter 10]{EllMaaMac12}. A proper exposition is too
lengthy to include here: I shall mention only the significance of expansion
in the results.

The role of the expansion rate in the evolution of perturbations
arises because when perturbations have a length scale much greater
than the contemporaneous Hubble scale, $c/H$, they are ``frozen in''.
This allows long wavelength perturbations to be tracked readily
through the transitions from inflation through radiation domination to
matter domination. (The Hubble scale has often been referred to in
this context as the Hubble horizon, although it does not agree with
any causal horizon. In a Big Bang the particle horizon at time $t_o$ is
at comoving radial coordinate distance
\[ u = \int_0^{t_o} \frac{\dx t}{a(t)}\]
 which in the de Sitter metric in
its expanding form, \eqref{deSit2}, is bounded above by $1/H$. Although
\eqref{deSit2} is not really a Big Bang model, since it contains no
matter, it appeared as the
approximate metric for the early universe in the initial 
models of inflation, and this may be why the Hubble scale became
called a horizon.)

As discussed in \citeasnoun{Dur15}, while the baryons and photons are
tightly coupled the whole content undergoes acoustic oscillations
under the competing effects of gravitational collapse and radiation
pressure. These resulting waves travel at the sound speed of this
medium, but become frozen in at decoupling, whence the peaks and
troughs observed in the CMB power spectrum.

The perturbation theory gives excellent agreement with the CMB
fluctuations, but those observations only test scales above 150 Mpc. It
is clear that nonlinear effects become important at smaller
scales. Theories of galaxy formation and early evolution begin with
collapse of a gas cloud, like star formation only larger, and then
proceed by accretion and merger. The many theoretical inputs have been
tabulated by \citeasnoun{Sco11}.

The main tool used to investigate the nonlinear phase of structure
formation, beyond nonlinear perturbation theory, is large Newtonian
N-body simulations. These can produce the principal features of the
actual distributions of galaxies, walls, filaments, voids
etc. However, they are not in perfect agreement with observation
quantitatively and have a shaky theoretical basis, since they are
non-relativistic. Their results do, for example, agree with the
relativistic perturbation theory of FLRW models in the conformal
Newtonian gauge \cite{ChiZal11}, but a recent paper has discussed a
situation where inhomogeneities that are ``easy to describe using the
linearized general relativity'' lead to a model that ``taken as a
whole lies in fact in the nonlinear regime'' \cite{Kor14}.

In a $\Lambda$CDM model, the first structures to form
are small, suggesting that massive galaxies form by multiple mergers.
It is possible to use a ``fossil
record'' of space- and time-resolved star formation histories,
and so show that massive galaxies grow their mass from the inside out. The
spheroidal parts seem to grow mainly about 5-7 Gyr ago. Earlier galaxies
do not have the typical virialized structures we see today (ellipticals
or spirals) and have now been seen (using the Hubble Space Telescope)
to undergo the mergers expected. Each merger appears to take about 0.5
Gyr and a massive galaxy will have undergone 4-5 by today.

There is growing evidence of correlations between the masses and sizes
of galaxies (in particular their spheroidal parts) and the masses of
their central supermassive black holes (SMBH); see for example
\citeasnoun{KorHo13}. How the SMBH controls the growth of a galaxy, given
that it is typically only a thousandth of the total mass, is not yet
entirely clear: it may be via the jets produced by the SMBH, fuelled
by matter from an accretion disk, heating the surrounding gas and so
controlling its collapse. What is clear is that this is highly
nonlinear and another important cosmological application of general
relativity.

\subsection{Other models}

Models other than FLRW can provide important cosmological information
in several ways. They test whether features of FLRW models are
peculiar to those models, and whether the models are robust under perturbations
of parameters, and they admit a wide variety of potentially observable new
effects. They may allow the nonlinear modelling of structures at a
level which FLRW perturbation theory cannot address.
 
The spatially homogeneous but anisotropic models, especially the
expanding Big Bang cases, were extensively investigated from the 1960s
onwards. They can be classified into the Bianchi types by their
symmetry groups. The singularities that occur can be quite
complicated, showing the oscillatory behaviour found by
\citeasnoun{Mis69}, and the more recently discovered ``Mussel
attractor'' \cite{ColHer05}. Over time, good choices of variables have been
found for the systems of ordinary differential equations that arise,
enabling very detailed and full studies
\cite{WaiEll97,EllMaaMac12}. Most of the dynamics can be understood
qualitatively by patching together segments of the evolutions of the
Bianchi I and II examples.

Two areas of particular interest have been the approach to the
singularity and the behaviour as $t\rightarrow \infty$. These are
relevant to our understanding of the real Universe inasmuch as there
are Bianchi models arbitrarily close to FLRW models for any given
accuracy of approximation and length of time, and those models can
differ radically from FLRW models at early and late times. However, a
positive cosmological constant tends to isotropize all Bianchi models
as $t\rightarrow\infty$ \cite{Wal83}. Nucleosynthesis, the CMB
observations, inflation, horizons and various quantum gravity theories
have also been studied in Bianchi models.

Among inhomogeneous models, the LTB models, the more general spherically
symmetric models which were also first considered by {\Lemaitre}, and
the Szekeres models \cite{Sze75}, have been the most extensively
explored. Inhomogeneous spacetimes can model over- and
under-densities, nonlinear gravitational waves, and anisotropic and
inhomogeneous initial conditions, and some fit the SN1a and CMB data
surprisingly well (for examples, see \citeasnoun[chapter 19]{EllMaaMac12}),
showing one should not too readily adopt the FLRW
explanations. Inhomogeneities may help explain the apparent
acceleration, as described below, but they also provide ways of making
detailed nonlinear models of localized structures, such as the Local
Group, M87, the Great Attractor and so on \cite{BolKraHel10}. In
modeling voids and clusters, it was found that velocity perturbations
were more effective in producing structure than the usual density
perturbations alone.

It is sometimes said that inflation can explain the observed
homogeneity and isotropy, despite the fact that almost all
inflationary models start by assuming it, at least for the observable
region. Calculations in specific models (summarized in \citeasnoun[chapters
  18--19]{EllMaaMac12}) suggest this is not so, and more needs to be
done by combining non-FLRW geometries and varying forms of inflation
to determine the true position. So far it seems that anisotropy and
inhomogeneity may suppress inflation, though there is also a model in
which averaged inhomogeneities act as the inflaton \cite{BucOba11}.

\section{Open questions, and possible future developments}

The standard model as we now have it has three big obvious unknowns in
the natures of the inflaton, the dark matter and the dark
energy. There are a number of terrestrial dark matter searches in
progress and it is possible one of them will resolve that issue. I do
not know of current experiments that could lead us to a fuller
understanding of the inflaton: one would like to know in more detail
how it governs the early universe dynamics and the generation of
fluctuations, and the interactions by which its energy-momentum is
converted to present-day matter.

For the apparent acceleration of the expansion, a number of possible
explanations have been put forward. While there may be astrophysical
effects requiring amendment of our calibrations of the supernovae, and
our understanding of absorption between them and us, which could alter
the inferred $m-z$ relation, this seems increasingly unlikely as
other, albeit still model-dependent, evidence for acceleration (from
BAO or gamma ray bursters, for example) accumulates. Confidence in
FLRW models, in which the various ways to define acceleration
\cite{BolAnd08} agree due to the symmetry, thus leads to some form of
dark energy being necessary.

There are four possible causes of the apparent acceleration other than
the simple cosmological constant currently used in modeling. One is a
previously unknown quantum field (``quintessence'') with some time
dependence: observations currently under way aim at limiting the
possible time development of such fields (assuming an FLRW
model). Both large and small scale anisotropies might provide
explanations, the former by using a non-FLRW geometry and the latter
both by effects on light propagation and by the backreaction. Lastly,
since the inferences depend on the relativistic FLRW models, a
modified gravity theory may be indicated. Of these the most popular
explanations are the first and last\footnote{As shown by
  M. Fairbairn's count that of the 591 papers submitted to the online
  arXiv in 2012 discussing dark energy, 287 concerned modified gravity
  theories.}.

Small scale anisotropies do affect measurements of $m-z$ both by their
effect on light propagation and by the backreaction corrections to
\eqref{eq:11.1}. However, the effects on light propagation, e.g.\ on
galaxy number counts \cite{BerMaaCla14}, are probably only at the
levels of accuracy claimed by ``precision cosmology'' \cite[chapter
  15]{EllMaaMac12}. The possible models of backreaction by small scale
inhomogeneities were put in doubt by the work of
\citeasnoun{GreWal11}, but may still provide a possible, and to me
appealing, explanation of the apparent acceleration
\cite{RouOstBuc13,Kor14}.

Large scale inhomogeneities could lead to an apparent acceleration.
Since observations cannot directly separate spatial from temporal
variations, a spatial variation could account for the observations,
and a number of models on these lines have been devised (see
\citeasnoun[chapter 15]{EllMaaMac12} for a review). However, it is not
easy to fit all the phenomena, especially those which combine data
from various $z$. For example, \citeasnoun{BulCliFer12} show that LTB models
cannot simultaneously explain the SN1a results, the BAO, the local
value of $H$ and the kinematic Sunyaev-Zel'dovich effect that arises
(see \citeasnoun{Bir99}) when the observed galaxies are in motion
relative to a frame in which the CMB is isotropic.

There are a great many projects under way to refine the present data
from the CMB, SN1a, BAO, lensing, kinematic Sunyaev-Zeldovich effect, and
other sources already described above. For example, there are at least
9 aimed at constraining the equation of state of dark energy (is it
the $w=-1$ of the cosmological constant?), there are a number of
experiments aimed at measuring B-mode polarizations and there are
several terrestrial dark matter searches. Most of these involve very
delicate measurement: for example, a 10\% difference in $w$ from $-1$
implies a change of only 0.04 magnitudes in an SN1a at $z=0.6$.

One further new window related to general relativity may be provided
by gravitational waves (see the remarks in \citeasnoun{Dur15} on the
BICEP2 results). The best current evidence for the existence of such
waves is provided by the very detailed measurements of the binary and
double pulsars, where the period changes agree well with the expected
energy loss through gravitational radiation. In the near future, the
ground-based laser interferometric detectors, with their recently improved
sensitivity, will give interesting results whether they make a
detection or not. (Since the predicted emission of expected sources
should be detectable, not seeing anything would cause a re-evaluation
of our theoretical understanding.) Pulsar timing arrays \cite{McL14}
are developing to the point of being very effective ways to detect low
frequency waves. There is still hope that the space-based
interferometric detector LISA will fly within some of the
readers' lifetimes.

In the seminar I referred to in the introduction, Sciama explained how
the 1960s data favoured the Big Bang theory, and described the
relevant FLRW models. We now have a great deal more than one known
piece of cosmological data, though it is perhaps disappointing that
that piece, i.e.\ the value of $H$, is still rather imprecisely known,
as the divergent values at the end of section \ref{Hmeasure}
show. Even more can be confidently anticipated. However, it is far
from clear whether or when the big open questions just mentioned will
be settled. It could even be that evidence emerges forcing us to
replace general relativistic dynamics for the Universe. What seems
likely is that expansion will continue to play a big role in our
models.

\section*{Acknowledgements} I am indebted to the historical articles
cited early in this review for many important references, to Thomas
Buchert, Ruth Durrer and an anonymous referee for very helpful
comments on an earlier draft which included saving me from a few
serious mistakes, to Piotr Flin, Michael Heller, Michael
Rowan-Robinson, Ian Roxburgh and Will Sutherland for various information about
past and present astronomy and astrophysics, and to Chris Blake and
Guy Pooley for allowing me to re-use their figures.

\section*{References}


\end{document}